\documentclass[11pt,a4paper]{article}
\usepackage{jheppub}
\usepackage{subfigure}
\usepackage{epic}

\newcommand{\beq}{\begin{eqnarray}}
\newcommand{\eeq}{\end{eqnarray}}
\newcommand{\non}{\nonumber\\}

\newcommand{\p}{\partial}
\newcommand{\Tr}{{\rm Tr}}

\newcommand{\cb}{\makebox(0,-2){$\bullet$}}

\newtheorem{theorem}{Theorem}
\newtheorem{corollary}{Corollary}

\title{Time-dependent stabilization in AdS/CFT} 

\author{Roberto Auzzi,}
\author{Shmuel Elitzur,}
\author{Sven Bjarke Gudnason}
\author{and Eliezer Rabinovici}

\affiliation{Racah Institute of Physics, The Hebrew University,
 Jerusalem 91904, Israel}

\emailAdd{auzzi(at)phys.huji.ac.il}
\emailAdd{elitzur(at)vms.huji.ac.il}
\emailAdd{gudnason(at)phys.huji.ac.il}
\emailAdd{eliezer(at)vms.huji.ac.il}

\abstract{

We consider theories with time-dependent Hamiltonians 
which alternate between being bounded and unbounded from below.
For appropriate frequencies dynamical stabilization can occur 
rendering the effective potential of the system stable. 
We first study a free field theory on a torus with a time-dependent
mass term, finding that the stability regions are described in terms
of the phase diagram of the Mathieu equation. 
Using number theory we have found a compactification scheme such as to
avoid resonances for all momentum modes in the theory. 
We further consider the gravity dual of a conformal field theory on a 
sphere in three spacetime dimensions, deformed by a doubletrace
operator. The gravity dual of the theory with a constant unbounded
potential develops big crunch singularities;
we study when such singularities can be cured by dynamical stabilization.
We numerically solve the Einstein-scalar equations of motion in
the case of a time-dependent doubletrace deformation and find that for 
sufficiently high frequencies the theory is dynamically stabilized and
big crunches get screened by black hole horizons. }

\begin{document}
\maketitle

\section{Introduction}

One of the main challenges facing string theory is to uncover what is
to actually become of various types of classical gravitational
singularities. The AdS/CFT framework is a particularly useful setup
for such a study.
On the one hand it provides complete non-perturbative information and
on the other hand, decays from a false vacuum in AdS \cite{CdL}
generically result classically in a big crunch. 
This was studied in
\cite{HH1,HH2,Craps,Elitzur:2005kz,Elitzur:2007zz,Craps:2009qc,Bernamonti:2009dp,de Haro:2006nv,Barbon:2010gn,Barbon:2011ta,Harlow:2010az,Maldacena:2010un}; in
particular it was argued in \cite{Barbon:2010gn,Barbon:2011ta} that
there are cases were the singular nature of a crunch is a question of
observables and conformal frame used to describe them. The
description of a big crunch as a boundary theory can be singular in
one frame and non-singular in another. In the bulk the singularity is
not resolved and yet infinite entropy is encoded on the
boundary. In one frame the Hamiltonian is always bounded and in the
other it is bounded as long as it exists. There are other more radical 
crunches which involve infinite energies and unbounded Hamiltonians on
the boundary
\cite{HH1,HH2,Elitzur:2005kz,Elitzur:2007zz,Barbon:2010gn,Barbon:2011ta}. 
In those cases there is no clear expectation that the crunch would or
should be cured. 
In this work we analyze boundary and bulk behavior in a class of cases
in which the boundary Hamiltonian's properties are somewhat midway
between the bounded and unbounded cases. 
We study here a class of time-dependent Hamiltonians which are unbounded
"half" of an infinite time and bounded during the other "half" of
time. This in an oscillatory manner. A butterfly flapping its wings. 
Such systems are interesting to discuss in their own right. In some
range of parameters it turns out that the energy of such systems is
effectively bounded from below. We will survey how the stability or
instability features of these boundary potentials manifest themselves
in the bulk. When crunches would occur, would black holes form to
shield the singularities?
In particular, one could pose the question if it is possible to turn
the mentioned big crunch into a near-crunch by means of such a
dynamical stabilization. 
As a prototype example we will consider in this paper the same
unbounded potential, however, multiplied by an oscillating function
such that half of the time the potential is bounded and half it is
not. 
Concretely we will consider an unbounded potential of the dual
boundary theory, oscillating with frequency $\omega$ as follows
\beq
V(\phi,t) = - V_0(\phi) \cos(\omega t) \, ,
\eeq
where $\phi$ is the field in the problem at hand.
An example of dynamical stabilization is Kapitza's pendulum
\cite{Kapitza} which is stabilized for frequencies above some critical 
frequency $\omega>\omega_c$. This was done using the method of
separation of time scales. 

In order to acquire some intuition we will first consider the example
of a free field theory. For the free field theory it will prove
convenient to break down the problem to the zero mode and higher
momentum modes. This leads us to study analogous quantum mechanics
problems before considering the field theory cases at hand.
We find in the case of a free massive field theory, when compactified
on a torus, that it will indeed be stabilized when the (angular)
frequency is larger than an order one number times the inverse
compactification radius. There is though an unwanted side effect
coming about by means of our treatment, i.e.~dynamically stabilizing
the system with oscillatory behavior, will in general lead to 
a secondary effect of resonances. This is a different type of
instability not present in the initially static unbounded
potential. An analysis in terms of number theory shows that in the
case of the free field theory it is possible to avoid the resonances
by choosing certain values of the compactification radius. 

The free massive field theory has the Mathieu equation as its equation
of motion, which shows up many places in time-dependent physical
problems. Tacoma bridge and the Paul trap are classic
examples. Other more recent phenomena involving the Mathieu equation
include reheating in the context of cosmology
\cite{Traschen:1990sw,Dolgov:1989us,Kofman:1994rk,Kofman:1997yn},
relativistic ion traps \cite{Durin:2003gj} as well as trapping of
particles using lasers \cite{Rahav:2003c}. 

In the interacting field theory the story is somewhat more
complicated and we do not have concrete proofs at hand. Intuition and
the method of separation of time scales provide evidence for the same
to occur in this case as well. A difference between the free and the
interacting theory is the expected thermalization in the latter. 
However for sufficiently high frequencies we have indications from the
zero mode by means of the quantum mechanics problem as well as from
the bulk that the boundary theory is dynamically stabilized. 

Concretely we solve the Einstein-scalar equations numerically in
global AdS${}_4$ deformed by a time-dependent doubletrace operator
finding that the big crunch may or may not appear depending on the
choice of frequency. In all cases we start by having a smooth initial
field configuration given by a Coleman-de Luccia (CdL) instanton
\cite{CdL} (restricted to the equator).

Most of the numerical studies in AdS space concerning Einstein-scalar
equations so far have focused on a massless scalar field
(e.g.~\cite{Pretorius,Bizon,Bizon2,GPZ,GPZ2}).
In order to be able to introduce a (non-irrelevant) multitrace
deformation \cite{Witten:2001ua,BSS,SS} in the case of AdS${}_{d+1}$,
it is required that the mass squared of the scalar obeys
\beq
-\frac{d^2}{4} \leq  m^2 < -\frac{d^2}{4} + 1 \, , \label{ranger}
\eeq
such that both possible fall offs of the scalar field
\beq
\phi \rightarrow \frac{\alpha}{r^{\Delta_-}} +
\frac{\beta}{r^{\Delta_+}} \, , \qquad
\Delta_{\pm}=\frac{d}{2} \pm \sqrt{\frac{d^2}{4} +m^2} \, ,
\eeq
are normalizable \cite{KW}.
The multitrace deformation is then realized as a boundary condition
relating $\alpha$ and $\beta$. Hence it proves crucial to
consider an AdS field with a mass squared in the range (\ref{ranger})
in order to allow the introduction of an external time-dependent
potential. 
A doubletrace deformation in the boundary theory
(which is proportional to $H(t)\mathcal{O}^2/2$, where $\mathcal{O}$
is an operator dual to the bulk field $\phi$)
corresponds to the boundary condition $\beta=H(t)\alpha$.
We numerically solve the Einstein-scalar equations subject to the
latter boundary condition with $H(t)=h_0\cos(\omega t)$.
This enables us to monitor the circumstances under which black holes
are formed. We are particularly interested in the $\omega$
dependence of the formation. 

Having set the scene we will start out by recalling the method of
separation of time scales classically (Sec.~\ref{Sect:classical}) and
quantum mechanically (Sec.~\ref{Sect:QM}) which we generalize to field 
theory in Sec.~\ref{Sect:free QFT} for both a free massive field
theory as well as for a quartically interacting theory. 
In Sec.~\ref{Sect:AdS} we study in detail the bulk manifestations of
the stabilities and instabilities on the boundary. We will present
evidence for big crunches which in some cases are indeed shielded
by black hole horizons. This provides evidence that 
thermalization indeed occurs and is responsible for the dynamical
stabilization in the case of interest. Specifically we will consider a
system in global AdS${}_4$ being dual to a 4D CFT on a sphere deformed
by a doubletrace operator.

\section{Classical mechanics} \label{Sect:classical}

In this section we will review the merits of the method of
separation of time scales. This method was used by Kapitza
\cite{Kapitza} to understand the dynamical stability of a vertically 
rapidly oscillating suspension holding a pendulum. The system is
stabilized with the pendulum pointing straight upwards due to the
kinetic energy of the very fast oscillations. Whenever the pendulum is 
slightly displaced from the point of equilibrium the effective
potential pushes the pendulum back into the straight upwards position
with damped slow oscillations.

Here we review the derivation of the effective Hamiltonian
(viz.~the Hamiltonian describing the drift part of the system) of a
rapidly oscillating potential using the method of separation of time
scales following \cite{LL,Rahav:2003a,Rahav:2003b,Rahav:2004} but
specialized to the Hamiltonian 
\beq
H = \frac{p^2}{2m} - V_0(x) \cos(\omega t) \, .
\label{eq:Horg}
\eeq
We decompose the particle $x(t)$ into a slowly moving part $X$ and
a rapidly moving part $\xi$ as
\beq
x(t) = X(t) + \xi(X,\dot{X},\omega t) \, , \qquad
\xi(X,\dot{X},\omega t) = \sum_{i=1}^\infty \omega^{-i} 
  \xi_i(X,\dot{X},\omega t) \, ,
\eeq
where $\xi$ is periodic with vanishing average in $\omega t$ and the
functions $\xi_i$ are chosen such that $X$ is independent of $\omega
t$. From the equation of motion
\beq
m\big(\ddot{X} + \ddot{\xi}\big) = V_0'(X+\xi) \cos(\omega t) \, ,
\label{eq:mosots_eom}
\eeq
we can use the chain rule on derivatives of $\xi$ and in turn
determine $\xi_i$ order by order in $1/\omega$. Doing so up till
order $\omega^{-4}$, it is possible to determine the drifting part of
the equation of motion, i.e.~the terms that do not average to zero
and hence are not absorbed into the $\xi_i$s. 
Expanding out the left hand side of \eqref{eq:mosots_eom} does not
give any non-vanishing terms upon time averaging (by definition),
hence we need to expand the right hand side
\beq
V_0'(X)\cos(\omega t) + \xi V_0''(X)\cos(\omega t) + \frac{1}{2}\xi^2
V_0^{(3)}(X)\cos(\omega t) + \cdots 
\eeq
The non-vanishing terms to 4th order upon time averaging are
\beq
\frac{1}{\omega^2}\overline{\xi_2 V_0''(X) \cos(\omega t)} 
+ \frac{1}{\omega^4}\overline{\xi_4 V_0''(X) \cos(\omega t)} \, , 
\eeq
where the bar denotes average with respect to time: $\overline{x}
\equiv \int_0^{2\pi} d(\omega t)x/(2\pi)$.
Calculating the coefficients of $\xi$
\begin{align}
\xi &= -\frac{V_0'\cos(\omega t)}{m\omega^2}
+\frac{2V_0''\dot{X}\sin(\omega t)}{m\omega^3}
+\frac{V_0'V_0'' \cos(2\omega t)}{8m^2\omega^4}
+\frac{3V_0^{(3)}\dot{X}^2\cos(\omega t)}{m\omega^4}
+ \mathcal{O}(\omega^{-5}) \, ,
\end{align}
we obtain the drifting part of the equation of motion 
\beq
m\ddot{X} = - \frac{1}{2m\omega^2}V_0' V_0''
+ \frac{3}{2m\omega^4} V_0'' V_0^{(3)} \dot{X}^2 
+ \mathcal{O}(\omega^{-5}) \, .
\eeq
This equation of motion can be obtained from the following effective
Hamiltonian 
\beq
H_{\rm eff} = \frac{P^2}{2m}
+ \frac{1}{4m\omega^2} [V_0']^2
+ \frac{3}{4m^3\omega^4} [V_0'']^2 P^2 
+ \mathcal{O}(\omega^{-5}) \, , \label{eq:Heff}
\eeq
with $P$ being the momentum conjugate of $X$. 
Physically we can explain this effective Hamiltonian as follows. Since
our choice of potential averages out to zero in time, to leading order
the theory is free. 
The correction to the potential equals  
$\frac{1}{2}m\overline{\dot{\xi}^2}$ up to 4th order in $1/\omega$, 
i.e.~it is the kinetic energy of the rapid oscillations
\cite{LL}\footnote{At the 4th order in $1/\omega$,
  $\frac{1}{2}m\overline{\dot{\xi}^2}$ gives the effective potential
  only up to a total derivative.}. This leading order potential is
\emph{always} confining.  
Notice that the dynamics of the drift degree of freedom $X$ happens to 
be conservative in spite of the fact that the original system has 
an explicit time-dependent Hamiltonian \eqref{eq:Horg}. 
For an alternative derivation of the effective potential and an
explanation of the effective conservation of energy, see
app.~\ref{app:FTderiv}. 


\subsection{Parameter counting}

Considering a monomial of the coordinate in a one dimensional
classical mechanics system
\beq
H = \frac{p^2}{2m} - \lambda^{n+1} x^n \cos(\omega t) \, ,
\eeq
we can infer that $\lambda$ has dimension of mass. Performing a
rescaling we can write
\begin{equation}
H = \Lambda \left[\frac{\tilde{p}^2}{2} - 
  q \,\tilde{x}^n \cos(2\tau)\right] \, , \qquad 
\tilde{p} \equiv \frac{d\tilde{x}}{d\tau} \, , \qquad
\Lambda \equiv \frac{1}{4}m\omega^2L^2 \, , \qquad
q \equiv \frac{4\lambda^{n+1}L^{n-2}}{m\omega^2} \,
, \label{eq:Hrescaled} 
\end{equation}
where $\Lambda$ (with unit of mass) sets the overall scale of the
problem and $q$ is a dimensionless parameter determining the rapidness
of the oscillations (small $q$ corresponds to fast
oscillations). Finally $\tilde{x}\equiv x/L$ and $\tau\equiv\omega
t/2$ are dimensionless spatial and time coordinates,
respectively. This shows that the system at hand has two parameters to
dial (for fixed boundary conditions). It is not possible to use
$\Lambda$ as an expansion parameter while it is indeed possible to
expand in small $q$ -- it is exactly the $1/\omega$ expansion.

\subsection{Quadratic potential}

Let us  consider the $n=2$ case of the Hamiltonian
\eqref{eq:Hrescaled}, i.e.~the case of the quadratic
potential. Applying the general expansion, the effective Hamiltonian
using eq.~\eqref{eq:Heff} for $V_0 = \lambda^3 x^2$ reads
\beq
H_{\rm eff} = \frac{P^2}{2m}
+ \frac{\lambda^6}{m\omega^2} X^2
+ \frac{3\lambda^6}{m^3\omega^4} P^2 
+ \mathcal{O}(\omega^{-5})\, . 
\label{eq:Heffx^2}
\eeq
The last term can be interpreted as modifying the effective mass with
respect to the bare mass due to the oscillations. 
The potential under consideration is very special though. In fact it
is not needed to make any expansion as the classical problem is
completely solvable. The equation of motion coming from
eq.~\eqref{eq:Hrescaled} with $n=2$ (dropping the tilde) reads
\beq
\ddot{x} - 2q \cos(2\tau) \, x = 0 \, ,
\eeq
which is the Mathieu equation \eqref{eq:Mathieu} with $a=0$. Let us
dwell on this classical problem for a while.
The general solution to this linear ordinary differential equation is
given in terms of the so-called Floquet solution, which can be
expressed as 
\beq
F(\tau) = e^{i\nu \tau} P(\tau) \ ,
\eeq
where $P(\tau)$ is a periodic function with period $\pi$ and $\nu$ is
the characteristic exponent which is in general complex. Hence, the
existence of an imaginary part of the characteristic exponent implies
that the solution either blows up in the infinite future or the
infinite past. 
The general solution is given by a linear combination of $F(\tau)$ and 
$F(-\tau)$ and therefore, unless one of the coefficients is tuned to
zero, the solution will blow up in the future if the imaginary part of
$\nu$ is non-zero. 

\begin{figure}[tbp]
\begin{center}
\includegraphics[width=0.5\linewidth]{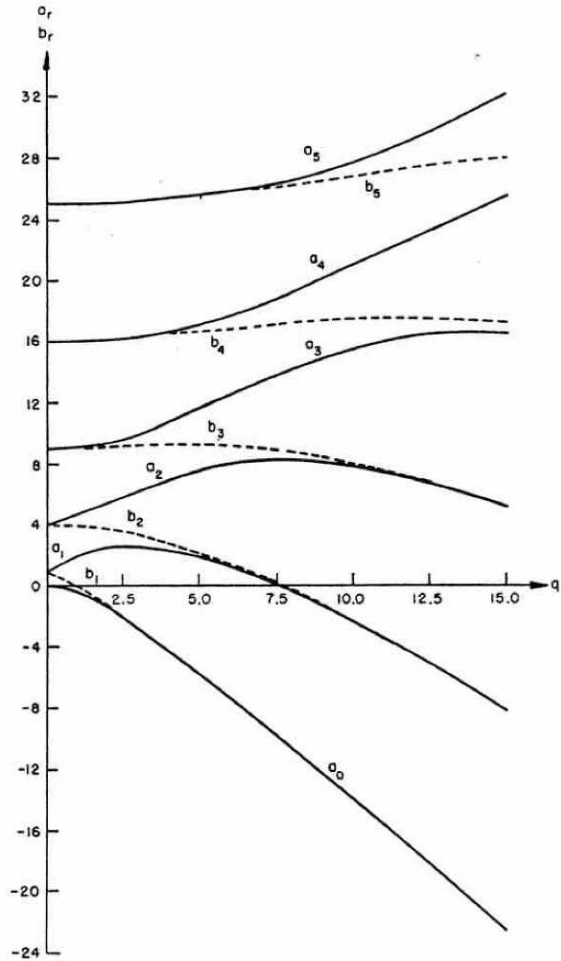}
\caption{Phase diagram of the Mathieu equation \eqref{eq:Mathieu}. 
  The stable regions are between the lines $a_i$ and $b_{i+1}$.
  For small $q\ll 1$ there are almost only stable regions except for
  thin resonance bands, while for large $q>a$ there are almost only
  unstable regions, except for thin stable bands. 
  In the unstable regions the characteristic exponent $\nu$ will have
  a non-zero imaginary part. This figure is taken from
  \cite{Abramowitz}. }
\label{fig:mathieu_phasediagram}
\end{center}
\end{figure}

Now we will first pose the question of stability. Let us define
stability by boundedness of the generalized 
coordinate $x$, i.e.~one could interpret it as the classical
``particle'' not running off to infinity exponentially fast. 
The answer in the classical case is given by the phase 
diagram of the Mathieu equation \cite{Abramowitz}, see
fig.~\ref{fig:mathieu_phasediagram}. The first few stable bands are
shown in table \ref{tab:stable_qbands}. 
\begin{table}[!htbp]
\begin{center}
\begin{tabular}{r||l|l}
band \# $s$ & $q_{s,{\rm start}}$ & $q_{s,{\rm end}}$ \\
\hline
\hline
$1$ & $0$ & $0.908046$ \\
$2$ & $7.51361$ & $7.57969$ \\
$3$ & $21.2986$ & $21.3032$ \\
$4$ & $41.9272$ & $41.9274$ \\
$5$ & $69.4284$ & $69.4284$ 
\end{tabular}
\caption{Stable bands in the variable $q$ of the Mathieu equation
  \eqref{eq:Mathieu} for $a=0$.} 
\label{tab:stable_qbands}
\end{center}
\end{table}
As compatible with physical intuition, the stable region is roughly
only given by small $q\lesssim 0.9$, except for very narrow bands in
$q$. According to eq.~\eqref{eq:Hrescaled} large frequencies
correspond to small $q$. Thus when the potential is oscillating very
rapidly, the forces practically cancel and we are left with a ``free
particle''. This agrees with intuition as well as the $1/\omega$
expansion. For an estimate of the amplitude of the particle position,
see app.~\ref{app:amplitude}.

\subsection{Quartic potential}

Turning to the $n=4$ case of the Hamiltonian \eqref{eq:Hrescaled} and
applying the expansion we obtain for $V_0 = \lambda^5 x^4$ the
effective Hamiltonian
\beq
H_{\rm eff} = \frac{P^2}{2m}
+ \frac{4\lambda^{10}}{m\omega^2} X^6
+ \frac{108\lambda^{10}}{m^3\omega^4} X^4 P^2 
+ \mathcal{O}(\omega^{-5})\, . 
\label{eq:Heffx^4}
\eeq
To our knowledge this case is not exactly solvable and hence we have
nothing to add to the results of the $1/\omega$ expansion, i.e.~the
effective potential is confining.

\section{Quantum mechanics}   \label{Sect:QM}

Following the work of Grozdanov-Rakovi\'c \cite{Grozdanov:1988} and
Rahav-Gilary-Fishman \cite{Rahav:2003a,Rahav:2003b} it is possible to
use the method of separation of time scales also to calculate the
quantum corrections to the Hamiltonian considered previously. 
This gives basically the effective Hamiltonian \eqref{eq:Heff} (with
appropriate ordering of the operators) as well as an extra quantum
correction (to 4th order). 
The spirit of the calculation is to find a unitary transformation of
the quantum Hamiltonian rendering it time independent. Again,
doing so exactly is a highly non-trivial task but order by order in 
$1/\omega$ this can be done systematically and the resulting effective
quantum Hamiltonian reads \cite{Rahav:2003a,Rahav:2003b}
\beq
H = \frac{p^2}{2m}
+\frac{1}{4m\omega^2} [V_0']^2
+\frac{3}{16m^3\omega^4} \left\{p,\left\{p, [V_0'']^2\right\}\right\}
+\frac{\hbar^2}{16m^3\omega^4} \big[V_0^{(3)}\big]^2
+\mathcal{O}(\omega^{-5}) \, ,
\eeq
where $x$ and $p$ are now operators. 
Let us apply this formula to two cases in turn in the following. 

\subsection{Quadratic potential}\label{sec:quantum_toy}
\begin{figure}[tbp]
\begin{center}
\includegraphics[width=0.6\linewidth]{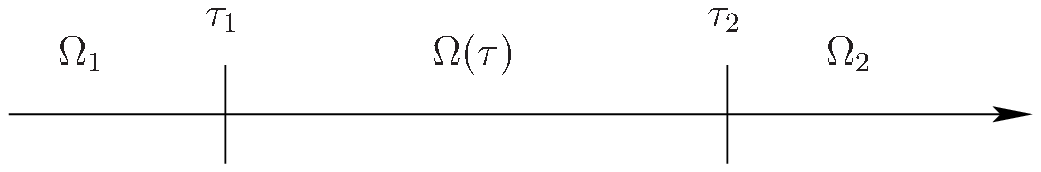}
\caption{The frequency of the time-dependent harmonic oscillator as
  function of time $\tau$. } 
\label{fig:time}
\end{center}
\end{figure}

In the case of $V_0=\lambda^3 x^2$ the effective quantum Hamiltonian 
reads
\beq
H = \frac{p^2}{2m} + \frac{\lambda^6}{m\omega^2}x^2 
+ \frac{3\lambda^6}{m^3\omega^4} p^2 
+\mathcal{O}(\omega^{-5}) \, .
\eeq
Note that the quantum Hamiltonian to 4th order in the $1/\omega$
expansion is exactly equal to its classical counterpart
(eq.~\eqref{eq:Heffx^2}) and hence also the quantum effective
potential is confining. 
As before this is a special case and we can say a lot more than
what is given by the $1/\omega$ expansion. In fact the problem can be
solved also in quantum mechanics. Lewis and Riesenfeld (LR)
\cite{Lewis:1968tm} found a method of invariants to solve
time-dependent harmonic oscillator problems. They considered the
generic Hamiltonian 
\beq
H = \frac{1}{2m} \left(p^2 + \Omega^2(\tau) \, x^2 \right)\, , 
\label{eq:LR_Hamiltonian}
\eeq 
with $\Omega^2(\tau) \in \mathbb{R}$ (not positive definite) and 
the corresponding classical equation of motion
\beq
\ddot{x} + \Omega^2(\tau) \, x = 0 \, , \qquad
\dot{x} = p \, . 
\label{eq:LR_classical_eom}
\eeq 
Let us make a lightening review of the results of the LR-method. First
we split time into three regions: a past ($\tau<\tau_1$) and a future
($\tau>\tau_2$) where the frequency of the harmonic oscillator is
constant; in between we turn on the time-dependent frequency
$\Omega(\tau)$, see fig.~\ref{fig:time}. If one desires, it is
possible to send $-\tau_1,\tau_2$ to infinity. 
This problem admits the construction of an invariant operator called
the Lewis invariant \cite{Lewis:1967}
\beq 
I = \frac{1}{2}\left[\frac{x^2}{\rho^2} 
+\left(\rho p - m \dot{\rho} x\right)^2 \right] \, , 
\eeq
which obeys Heisenberg's equation of motion while
\beq
m^2 \ddot{\rho} + \Omega^2(\tau) \rho - \frac{1}{\rho^3} = 0 \, , 
\label{eq:EP}
\eeq
is the Ermakov-Pinney equation. 
The eigenvalues of $I$ can be shown to be time independent
\cite{Lewis:1968tm}, while its eigenstates are time dependent. The
trick is that the eigenstates of the invariant operator $I$ can be
mapped to those of the Hamiltonian and hence the transition
probabilities for any given system can be calculated in terms of the
eigenstates of the original Hamiltonian \eqref{eq:LR_Hamiltonian}. The
general solution to eq.~\eqref{eq:EP} for 
constant $\Omega$ is \cite{Lewis:1967}
\beq
\rho(\tau) = \pm \sqrt{|\Omega|^{-1}
\left[\cosh\delta \pm \sinh\delta 
\sin\left(2\Omega \tau/m + \varphi\right)\right]} \ , 
\eeq
where $\delta,\varphi\in\mathbb{R}$ are real parameters. The
parameter $\delta$ encodes all information for calculating the
transition probabilities from any given state at time $\tau<\tau_1$
to any state at time $\tau_2$. Unfortunately, $\delta$ is not
analytically calculable, it can however be computed numerically. 

For our purposes, we do not really need to calculate the occupation of
particular states in detail, we merely would like to know if the
quantum system is bounded in the sense that the wave function yields a 
bounded expectation value of the position operator, or alternatively
that the average number of created particles remains bounded. 
The solution to the Ermakov-Pinney equation \eqref{eq:EP} can be
written as follows \cite{Pinney:1950,Leach:2008}
\beq
\rho(\tau) = \sqrt{a u^2(\tau /m) + c v^2(\tau /m) + 2b u(\tau
  /m)v(\tau /m)} \ ,  
\label{eq:EPsol}
\eeq
where $u(\tau /m),v(\tau /m)$ are two linearly independent solutions to
eq.~\eqref{eq:LR_classical_eom} and $ac - b^2 = 1/W^2$, where 
$W = u \dot{v} - v \dot{u}$ is the constant Wronskian of the solutions. 

The argument now goes as follows, if the classical system satisfies
our demand on boundedness, then the two solutions $u(\tau /m),v(\tau
/m)$ are necessarily bounded and hence for any set of finite constants 
$\{a,b,c\}$ implied by any finite boundary condition for the quantum 
auxiliary function $\rho(\tau)$, its behavior will also obey the
demand on boundedness. This establishes equality between boundedness
of the classical and the quantum systems.

\subsection{Quartic potential}
In this case, $V_0=\lambda^5 x^4$ and hence we have
\beq
H = \frac{p^2}{2m} + \frac{4\lambda^{10}}{m\omega^2} x^6
+\frac{27\lambda^{10}}{m^3\omega^4} \left\{p,\left\{p,x^4\right\}\right\}
+\frac{36\hbar^2\lambda^{10}}{m^3\omega^4} x^2
+\mathcal{O}(\omega^{-5}) \, .
\eeq
This theory cannot be solved analytically and we have to rely on the
assumption that the $1/\omega$ expansion is valid and equivalently the
effective quantum potential is confining. 

\section{Field theory}  \label{Sect:free QFT}

\subsection{Quadratic potential}

Consider the Lagrangian density 
\beq
\mathcal{L} = \frac{1}{2}(\p_\mu \phi)^2 
  + \frac{1}{2} m^2 \cos(\omega t) \, \phi^2 \, , 
\eeq
where $\phi(t,\vec{x})\in\mathbb{R}$ is a real-valued scalar field in 
$(d+1)$-dimensional flat spacetime, $m^2\in\mathbb{R}_{>0}$ and
$\omega$ is an externally given fixed frequency. 
For convenience we choose the field to be tachyonic at time $t=0$. 
The equation of motion is
\beq
\ddot{\phi} + \left[a - 2q\cos(2\tau)\right]\phi = 0 \, , \qquad
\phi = \phi(\tau,\vec{k}) \, , 
\label{eq:Mathieu}
\eeq
where we have Fourier transformed the field in the spatial directions
and rescaled time as $\tau\equiv\omega t/2$ while the double-dot in
eq.~\eqref{eq:Mathieu} denotes the second derivative with respect to
$\tau$. 
Eq.~\eqref{eq:Mathieu} is the Mathieu equation on canonical form
\cite{Mathieu}, while the parameters are related to the original
Lagrangian density as follows
\beq
a \equiv \frac{4 \vec{k}^2}{\omega^2} \, , \qquad
q \equiv \frac{2 m^2}{\omega^2} \, . 
\eeq 
The external frequency $\omega$ sets the scale of the problem, while
the parameter $q$ corresponds to the mass squared and finally the
parameter $a$ corresponds to the two-norm squared of the momentum
vector. $a$ is the parameter in the equation adding structure to the
system due to the quantum field theory as compared to the quantum
mechanics case.

There are two types of impinging instabilities threatening the
system. The first is the unboundedness of the potential which we have
chosen to cure by making the potential oscillate rapidly (i.e.~we
consider $q\ll 1$, see fig.~\ref{fig:mathieu_phasediagram}). The
second issue concerns resonant modes appearing upon this dynamical 
stabilization of the potential (i.e.~for $a$ taking on integer values
resonant modes appear, see fig.~\ref{fig:mathieu_phasediagram}). 
\emph{Let us emphasize that this instability is not that of the
 potential being negative but merely a secondary effect due to the
 oscillatory behavior of the potential providing opportunity for
 resonances in the system at hand}.
The critical momentum magnitude above which all momentum modes render
the potential of the theory stable is 
\beq
k_i^2 > k_{\rm critical}^2 = m^2 \, .
\label{eq:critical_momentum}
\eeq
Above this critical momentum only the resonances reside.
For a continuum of momentum modes, some modes are bound to hit the 
resonances, which suggests that if we want stability we should
compactify space. For simplicity we will consider only toroidal
compactification of the spatial dimensions.
Our spacetime manifold is therefore $\mathbb{R}\times T^d$ and we
denote the compactification radii as $R_i$, $i=1,\ldots,d$. Now the
momenta are given in terms of a set of integers $\{n_i\}$ as follows
\beq
k_i = \frac{n_i}{R_i} \, , 
\eeq
and hence 
\beq
a = \sum_{i=1}^d \theta_i n_i^2 \, , \qquad
\theta_i \equiv \frac{4}{\omega^2 R_i^2} \, .
\eeq
According to eq.~\eqref{eq:critical_momentum} the potential becomes
positive definite if we choose the compactification radii such that 
\beq
\frac{1}{R_i^2} > m^2 \, , \;\; \forall i = 1,\ldots, d \, . 
\eeq
This ensures that there is no instability due to the potential in the 
theory while it does not guarantee the absence of
resonances which can and do occur in general. 
In the remainder of this section we will study a compactification
scheme such as to avoid the resonant modes entirely.

Avoiding the resonant momentum modes corresponds to choosing the
parameters $\theta_i$ such that the quantized momentum modes hit only
stable zones of the phase diagram \ref{fig:mathieu_phasediagram}; this
means they should avoid hitting any integer squared $s^2$,
$s\in\mathbb{Z}_{>0}$. The size of the unstable band can be
estimated for large $s$ as \cite{Abramowitz} 
\beq
\Delta_s \sim \frac{q^s}{s^{s-1}} \, ,
\label{eq:size_instability_band}
\eeq
which also calls for very small $q\ll 1$. 
The stability condition now amounts to the following inequality
\beq
\left|\sum_{i=1}^d\theta_i n_i^2 - s^2\right| > \Delta_s \, , 
\eeq
which we will simplify by setting all the $\theta$s equal $\theta_i =
\theta$ (hence $R_i=R$) and thus
\beq
\left|\sum_{i=1}^d \frac{n_i^2}{s^2} - \frac{1}{\theta}\right| > 
\frac{e^{-s A}}{\theta \, s^{s+1}} \equiv \psi(s) \, , 
\label{eq:stability_condition}
\eeq
where $A \equiv -\log q > 0$. 

Instead of addressing this problem directly, let us consider an easier 
but as we shall see, more general problem, viz.~consider the stability
condition
\beq
\left|\frac{p}{q} - \frac{1}{\theta}\right| > \psi(q) \, ,
\label{eq:mainineq}
\eeq
where $p,q\in\mathbb{Z}_{>0}$ are integers. Now the strategy is to
find the best approximation $p/q$ to $1/\theta$ and try to estimate
how big their difference will be as function of $q$. If it will be big
enough, i.e.~bigger than $\psi(q)$, then we are guaranteed absence of
resonances. 

Let us define $\theta' \equiv 1/\theta$ and consider the Diophantine
approximation to $\theta'$
\beq
\left|\frac{p}{q} - \theta'\right| < \frac{1}{M q^2} \, ,  
\eeq
where $M$ is a constant of order one depending on $\theta'$. In
general $\theta'$ is a real number but we will consider the case where
it is an \emph{irrational} number,
$\theta'\in\mathbb{R}\backslash\mathbb{Q}$ and positive. 
It will prove convenient to utilize the continued fraction as a method
for obtaining the Diophantine approximation $p/q$ to the irrational
number $\theta'$. The continued fraction is given by
\begin{align}
\frac{p_n}{q_n} =
a_0 + \frac{1}{a_1 + \frac{1}{\begin{array}{lll}
a_2 + \\[-9pt] &\ddots\\[-9pt] && a_{n-1} + \frac{1}{a_n}\end{array}}}
\, ,
\end{align}
where $\{a_i\}$ are positive integers and $p_n/q_n$ is called the
$n$-th principal convergent, while it holds that 
\beq
\lim_{n\to\infty} \frac{p_n}{q_n} = \theta' \, . 
\eeq
Hence if $\theta'$ is rational a solution exists. On
the other hand, if $\theta'$ is irrational, there will only exist an 
arbitrarily good approximation in form of a rational number. The
question is how fast it converges, viz.~how good the rational number
$p/q$ approximates $\theta'$ as function of $q$. We need to invoke a
few theorems due to Euler and Lagrange in order to answer that
question. 
\begin{theorem} \cite{Lang,Cassels}
For $n\geq 0$ we have
\beq
q_n p_{n-1} - p_n q_{n-1} = (-1)^n \, . 
\eeq
\end{theorem}
\begin{corollary}\label{corollary} \cite{Lang,Cassels}
If $a_1, a_2, \ldots$ are positive integers, then $p_n$ and $q_n$ are
relatively prime and $0 < q_1 < q_2 < \cdots$ 
forms a strictly increasing sequence of integers. 
\end{corollary}
\begin{theorem} \cite{Lang,Cassels}
The best approximations to $\theta'$ are the principal convergents to
$\theta'$. In fact for $n\geq 1$, $q_n$ is the smallest integer
$q>q_{n-1}$ such that $||q \theta'||<||q_{n-1}\theta'||$.\footnote{We
  adopt the notation $||\xi||$ as the distance between $\xi$ and the
  nearest integer.}
\end{theorem}
In view of the above theorem and the fact that $\{q_n\}$ is a strictly 
increasing sequence of integers, we just need to determine how good
the continued fraction approximates $\theta'$ for a given $q_n$. 
\begin{theorem} \cite{Lang,Cassels}
For even $n$, the $n$-th principal convergents of $\theta'$ form a
strictly increasing sequence converging to $\theta'$. For odd $n$, the
$n$-th principal convergents of $\theta'$ form a strictly decreasing
sequence converging to $\theta'$. Furthermore, we have
\beq
\frac{1}{2q_{n+1}} < 
\frac{1}{q_{n+1} + q_n} < 
|q_n \theta' - p_n| < 
\frac{1}{q_{n+1}} \, . 
\eeq
\label{th:nprincipal}
\end{theorem}
It then follows from theorem \ref{th:nprincipal} that
\beq
\left|\theta' - \frac{p_n}{q_n}\right| > \frac{1}{2q_{n+1}q_n} \, , 
\label{eq:est}
\eeq
which gives us an estimate on how bad the approximation is iff we
know how much larger $q_{n+1}$ will be compared to $q_n$. 
It can be shown that 
\beq
q_{n+1} = a_{n+1}q_{n} + q_{n-1} \, , 
\label{eq:recursionofqs}
\eeq
holds. Using
corollary \ref{corollary} we can infer that
\beq
q_{n+1} < (a_{n+1}+1)q_n \, ,
\eeq
and hence we can write eq.~\eqref{eq:est} as
\beq
\left|\theta' - \frac{p_n}{q_n}\right| > \frac{1}{2(a_{n+1}+1)q_n^2}
\, .
\label{eq:est2}
\eeq
Comparing the two right-hand sides of eqs.~\eqref{eq:est2} and
\eqref{eq:mainineq} we arrive at the following condition 
\beq
a_{n+1}+1 < \frac{\theta}{2} q_n^{q_n-1} \exp(q_n A) \, , 
\label{eq:condition}
\eeq
which when satisfied guaranties absence of resonances. 

An example which satisfies the condition \eqref{eq:condition} is the
golden ratio: $1/\theta=\frac{1+\sqrt{5}}{2}$ which has the integers
$a_i=1, i=1,2,\ldots$ and is 
therefore the slowest converging irrational number. Using
eq.~\eqref{eq:recursionofqs} it is seen the $q_n$s form a Fibonacci
series and hence $q_{n}<q_{n+1}<2q_{n+1}$. Plugging the $a_i$s into
eq.~\eqref{eq:condition} the left-hand side is constant while the
right-hand side is rapidly growing with $n$. 
Another example is $1/\theta=e$ having cyclic $a_i$s: 
$[a_0,a_1,\cdots] = [2,1,2,1,1,4,1,1,6,1,1,8,1,1,10,\cdots]$. This
number also satisfies the inequality \eqref{eq:condition}. 
Square roots of integers always have a periodic series in $\{a_i\}$
and hence are easy to check if they satisfy eq.~\eqref{eq:condition}.

The condition \eqref{eq:condition} is valid for 
eq.~\eqref{eq:mainineq} which is a generalization of the problem we
want to solve, namely that of eq.~\eqref{eq:stability_condition}.
$\sum_{i=1}^d n_i^2/s^2$ is not an arbitrary rational number $p/q$ but
(for finite dimensions $d<\infty$) it is a subset of rational numbers,
so the condition \eqref{eq:condition} is stronger than what we
need. Thus if it is satisfied also eq.~\eqref{eq:stability_condition} is
satisfied and hence there will not exist any momentum mode in the
compactified field theory hitting the instability bands of the phase
diagram of the Mathieu equation (no resonant modes).

\subsection{Damage control}\label{sec:damage_control}

Let us consider how bad things go if one should not take our advise on
the suitable values of $\theta$. 
We will here consider the consequences in the case that the chosen
value of $\theta$ gives rise to one or more modes hitting
a resonance band. 
We can estimate the time $t_\Lambda$ it takes until the norm of the
unstable mode has reached the size $\Lambda$ as
\beq
t_\Lambda \sim \frac{2}{\omega|\Im(\nu)|} \log\Lambda =
\frac{R\sqrt{\theta}}{|\Im(\nu)|}\log\Lambda \, . 
\label{eq:lifetime}
\eeq
In addition we need an estimate of the size of the imaginary
part of the characteristic exponent $\nu$, which can be found in the
literature \cite{McLachlan}, valid for $q\ll a$ (an alternative method
for estimating $\nu$ can be found in app.~\ref{app:nu})
\beq
\Im(\nu) \sim \pm \frac{\sqrt{(a_s - a)(a - b_s)}}{2s} \, , 
\eeq
where $a$ is the parameter of the Mathieu equation, while $a_s$
($b_s$) is the upper (lower) boundary of the $s$-th instability
band. Let us assume the worst case scenario, i.e.~the mode hitting 
right in the middle of the instability band giving the maximal
imaginary part of $\nu$ in which case (for $q\ll 1$) $a = (a_s+b_s)/2$
and hence 
\beq
\Im(\nu_{\rm max}) \sim \frac{\Delta_s}{4s} = \frac{q^s}{4s^s}\, , 
\eeq
where $\Delta_s$ is the size of the $s$-th instability band
\eqref{eq:size_instability_band}. This value is a rapidly decreasing
function of $s$ and hence the ``lifetime'' in eq.~\eqref{eq:lifetime}
will be parametrically large for high enough $s$. 
Now that one can tolerate the large life times associated with large
values of $s$, one needs to remove the thread of resonances for 
smaller values of $s$. There are two tools available for that; an
appropriate choice of $\theta$ or a sufficiently small $R\omega$.

\subsection{Quantizing the compactified field theory}

We have now obtained a classical field theory which under certain
conditions is stable in the sense that for each momentum mode, the
equation of motion gives a -- from the other modes decoupled --
bounded wave function under time evolution. For a single time
dependent harmonic oscillator we have seen in
sec.~\ref{sec:quantum_toy} that the quantum state evolves in a
controlled manner basically determined by the classical behavior. 
Let us remark that $\Omega^2(\tau)$ of eq.~\eqref{eq:LR_classical_eom}
is allowed to contain a non-zero constant term $a$ which is appearing
in eq.~\eqref{eq:Mathieu}. 
The time evolution of the quantum field theory state can be explicitly
calculated \cite{Vergel:2009st} which is basically done by acting with
the Hamiltonian on a state vector living in a Schwartz space in the 
Schr\"odinger representation and finally composing an infinite
product of these over all momentum modes. 
With respect to our stability/boundedness criteria, we are basically
home free due to the fact that we are dealing with a free
(i.e.~non-interacting) field theory whose infinitely many harmonic
oscillators are mutually uncoupled. We would however like to address
the question of whether the average number of particles created
remains finite, summing over all momentum modes. For that we use the
solution \eqref{eq:EPsol} subject to the boundary conditions
$\rho(0)=(\theta n^2-2q)^{-1/4}$ and $\dot{\rho}(0)=0$ which in turn we
choose to feed with the Mathieu cosine and sine functions expanded in
small $q$ (see app.~\ref{app:amplitude}). For the characteristic
exponent $\nu$ we use the approximation \eqref{eq:nuexpansion}. 
In order to calculate the transition probability we need a quantum
mechanical integration constant $\delta$ calculated for each mode
$n$. The following integrated form of eq.~\eqref{eq:EP}
\cite{Lewis:1968tm} (for $m=1$)  
\beq
\dot{\rho}^2 + \Omega^2 \rho^2 + \frac{1}{\rho^2} =
2|\Omega|\cosh\delta \, , \qquad
\Omega^2 = \theta n^2 - 2 q \cos(2\tau) \, , 
\eeq
defines $\delta$. Plugging in the above described solution
$\rho$ we find
\begin{align}
\cosh\delta - 1 = & \, 
\frac{q^2}{4\theta^2 n^4\left(1-\theta n^2\right)^2}
\Big\{3+\theta n^2+(1-\theta n^2)\cos(4\tau) \\
&- 2(1+\sqrt{\theta}n)\cos\left[2(1-\sqrt{\theta}n)\tau\right]
- 2(1-\sqrt{\theta}n)\cos\left[2(1+\sqrt{\theta}n)\tau\right]\Big\}
+ \mathcal{O}(q^3) \, , \nonumber
\end{align}
with $\tau>0$.
The average number of particles $N$ \cite{Lewis:1968tm} is then
\beq
\langle N\rangle = \sum_{n_1,\ldots,n_d=1}^\infty
\frac{1}{2}\left(\cosh\delta -1\right) \, ,
\eeq
with $n\equiv \sqrt{n_1^2+\cdots+n_d^2}$. 
Expanding the above expression in large $n\gg 1$ we obtain the average
number of created particles
\beq
\langle N\rangle = \sum_{n_1,\ldots,n_d=1}^\infty
q^2 \left(\frac{1-\cos(4\tau)}{8\theta^3 n^6} 
+ \mathcal{O}(n^{-7})\right) + \mathcal{O}(q^3) \, ,
\eeq
from which it is seen that $\langle N\rangle$ converges for $d\leq 5$
spatial dimensions. 
Similarly we can calculate the total energy of the system, neglecting
the mass of the created particles, as
\beq
E_{\rm tot} \simeq \frac{1}{R} \sum_{n_1,\ldots,n_d=1}^\infty
\frac{n}{2}\left(\cosh\delta -1\right) \, ,
\eeq
which then converges for $d \leq 4$ spatial dimensions.

\subsection{Quartic potential}

For the field theory
\beq
\mathcal{L} = \frac{1}{2}(\partial_\mu \phi)^2 -
\frac{\lambda}{4} \cos(\omega t) \phi^4 \, ,
\eeq
the detailed information which was available for the quadratic
potential is not readily accessible. 
In the following we will take a leap of faith and assume that the
system thermalizes due to the fact that it is an interacting theory. 
The actual calculations we will perform will be related to the ABJM
theory \cite{ABJM} on a sphere. 
In the next section we consider the gravitational dual of such a
theory with time-dependent couplings. In the process we will provide
evidence from the bulk that the theory indeed thermalizes.

\section{AdS/CFT}  \label{Sect:AdS}

We will study the gravitational bulk manifestations in the
regions of (in-)stabilities obtained for the boundary field theory. 
Our expectations are that the gravitational dual of an unstable
boundary potential will result in a big crunch. For those values of
parameters for which the boundary theory is eventually stable, we
expect that a black hole (BH) horizon will form.
We will consider the setup of 11-dimensional supergravity in
AdS${}_4\times (S^7/\mathbb{Z}_k)$ as a ground on which to identify the
appropriate bulk field configurations, corresponding to the
different types of boundary potentials we study. 
$\mathcal{N}=8$ supergravity in four dimensions describes the
massless sector of the compactified 11-dimensional supergravity on
$S^7$. It is possible to introduce a consistent truncation involving
only gravity as well as a single scalar field $\phi$
\cite{Duff:1999gh}. We thus consider the following effective action 
\beq
S = \int d^4 x \, \sqrt{- g} \left(  
\frac{R}{2} -\frac{1}{2} \partial_\mu \phi \partial^\mu \phi - V(\phi)
\right) \, , \label{azio}
\eeq
with $V(\phi)=-2-\cosh (\sqrt{2} \phi)\approx -3 -\phi^2+ O(\phi^4)$. 
As pointed out in \cite{Craps:2009qc,Bernamonti:2009dp},
 the same consistent truncation can be used for the more general 
 compactification of 11-dimensional supergravity on the quotient space
 $S^7/\mathbb{Z}_k$.
The scalar field $\phi$ has a mass squared $m^2=-2$, which is above
the Breitenlohner-Freedman bound \cite{BF} $m_{\rm BF}^2=-9/4$.
The AdS metric in global coordinates can be written as
\beq
ds^2= -(1+r^2) \, dt^2 + \frac{dr^2}{1+r^2} +r^2 \, d \Omega_2^2 
=-\frac{dt^2}{\cos^2 x }+ \frac{dx^2}{\cos^2 x} +\tan^2 x \, d
\Omega_2^2 \, , 
\eeq
where $r=\tan x$.
The scalar field goes to zero asymptotically as
\beq
\phi(r) \approx \frac{\alpha}{r}+ \frac{\beta}{r^2} \, . \label{asintotico}
\eeq

According to the AdS/CFT correspondence, M-theory in global 
${\rm AdS}_4 \times (S^7/\mathbb{Z}_k)$ is dual to ABJM theory
\cite{ABJM} on $S^2 \times \mathbb{R}$, i.e.~a ${\rm U}(N)\times {\rm
  U}(N)$ superconformal Chern-Simons (CS) theory with matter and CS
levels $+k$ and $-k$. It contains four $\mathcal{N}=2$ chiral
superfields $Y_s$ transforming in the $(N,\bar{N})$ representation of
the gauge group. 
Due to the conformal coupling, the four scalars $Y_s$ acquire masses 
proportional to $1/R_{S^2}$ which we set to unity.
The field $\phi$ of the consistent truncation in eq.~(\ref{azio})
corresponds to the following scalar mass operator in ABJM theory
\beq
\mathcal{O}= \frac{1}{N} \Tr\left(Y_1 Y_1^\dag+Y_2 Y_2^\dag-Y_3 Y_3^\dag - Y_4 Y_4^\dag\right) \, .
\eeq 
Multitrace deformations \cite{Witten:2001ua,BSS} of the form
$S=S_0- N^2 W(\mathcal{O})$, correspond to the following boundary
condition
\beq
\beta =  W'(\alpha) \, .
\eeq
This is possible due to the fact that the mass of the scalar field
lies in the interval 
\beq
-\frac{9}{4} < m^2 < -\frac{5}{4} \, , \label{massaranger}
\eeq
and hence both scalar-wave fall offs of eq.~(\ref{asintotico}) are
normalizable.

By appropriately choosing $W(\mathcal{O})$, it is possible to
construct unbounded potentials in the boundary theory
for which $Y_s=0$ is a metastable vacuum;
for instance by choosing $W=H_3 \mathcal{O}^3/3$ with an arbitrary
sign for $H_3$, or alternatively $W=H_2 \mathcal{O}^2/2$ with
$H_2<0$.  The first case corresponds to AdS-invariant boundary
conditions \cite{Hertog:2004dr}. We first study the case of a
time-independent unstable potential and identify the bulk signature of
that and then we introduce time dependence \cite{HH1}.

\subsection{Instantons and crunches}

Let us first consider the case of a time-independent multitrace
operator. When the multitrace operator renders the boundary theory
metastable, this metastability manifests itself in the bulk as a
Coleman-de Luccia instanton \cite{CdL} when the multitrace operator is
marginal and by some other type of bubble when it is a relevant
operator \cite{Barbon:2010gn,Barbon:2011ta}.
In the AdS${}_4$ setting we are considering, instantons were studied
in detail in \cite{HH1,HH2}. The following spherically symmetric
Ansatz for the metric is used
\beq
d s^2 = \left(\frac{d\rho^2}{b^2(\rho)} + \rho^2 d \Omega_3^2 \right) \, .
\eeq 
The Euclidean action reads
\beq
S=\int d^4 x \, \sqrt{g} \left(  
-\frac{R}{2} +\frac{1}{2} \partial_\mu \phi \partial^\mu \phi + V(\phi) \right) 
- \int_{\partial} d ^3 x   \sqrt{h_c}  K +  \, S_{\rm ct} \,   +\, S_\partial \, ,  
\eeq
where $h_c$ is the induced metric on the cutoff surface,
$K$ is the trace of the extrinsic curvature of the boundary
(giving the Gibbons-Hawking term),
$S_{\rm ct}$ and $S_\partial$ contain counter terms and the boundary
terms, respectively. In the AdS${}_4$ setup with $m^2=-2$ which we
study, $S_{\rm ct}$ and $S_\partial$ take the forms
 \cite{Balasubramanian:1999re,Papadimitriou:2007sj}
\begin{equation}
S_{\rm ct} =  \int_{\partial} d ^3 x \, \sqrt{h_c} \left(
  2+ \frac{R(h_c)}{2} +\frac{1}{2} \Delta_- \phi^2  \right) \, , \qquad
S_\partial=
  \int_{\partial} d ^3 x \, \sqrt{h} \left(
  W(\alpha) -\alpha W'(\alpha)+ \alpha \beta  \right) \, , \nonumber
\end{equation}
where $\Delta_-=1$, $h$ is the induced metric on the boundary
and $R(h_c)$ is the Ricci scalar of the induced metric on the cutoff
surface. 

\begin{figure}[ht]
\begin{center}
\mbox{\subfigure[$\phi(x)$]{\epsfxsize=2.5in \epsffile{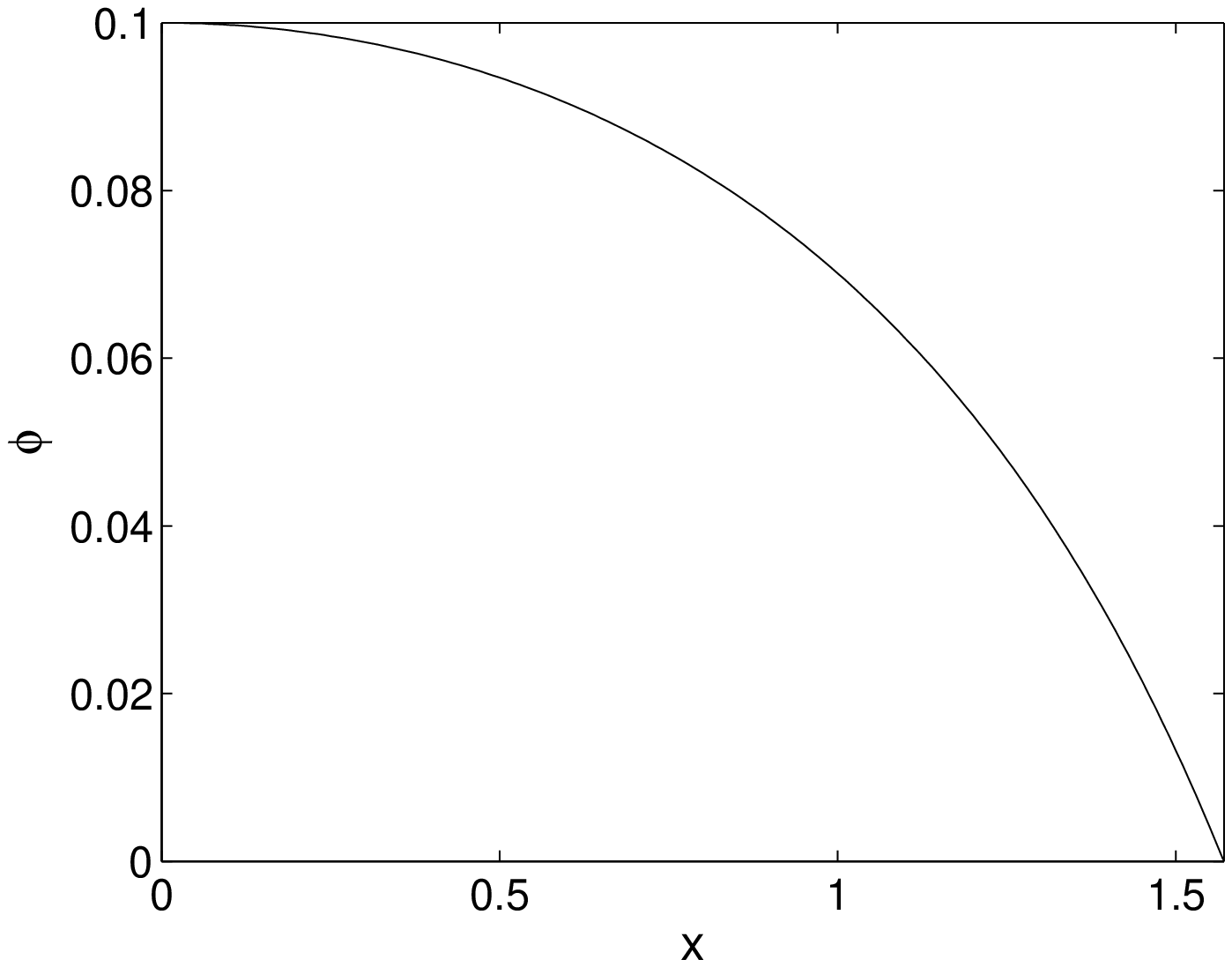}}\quad
\subfigure[$A(x)$]{\epsfxsize=2.5in \epsffile{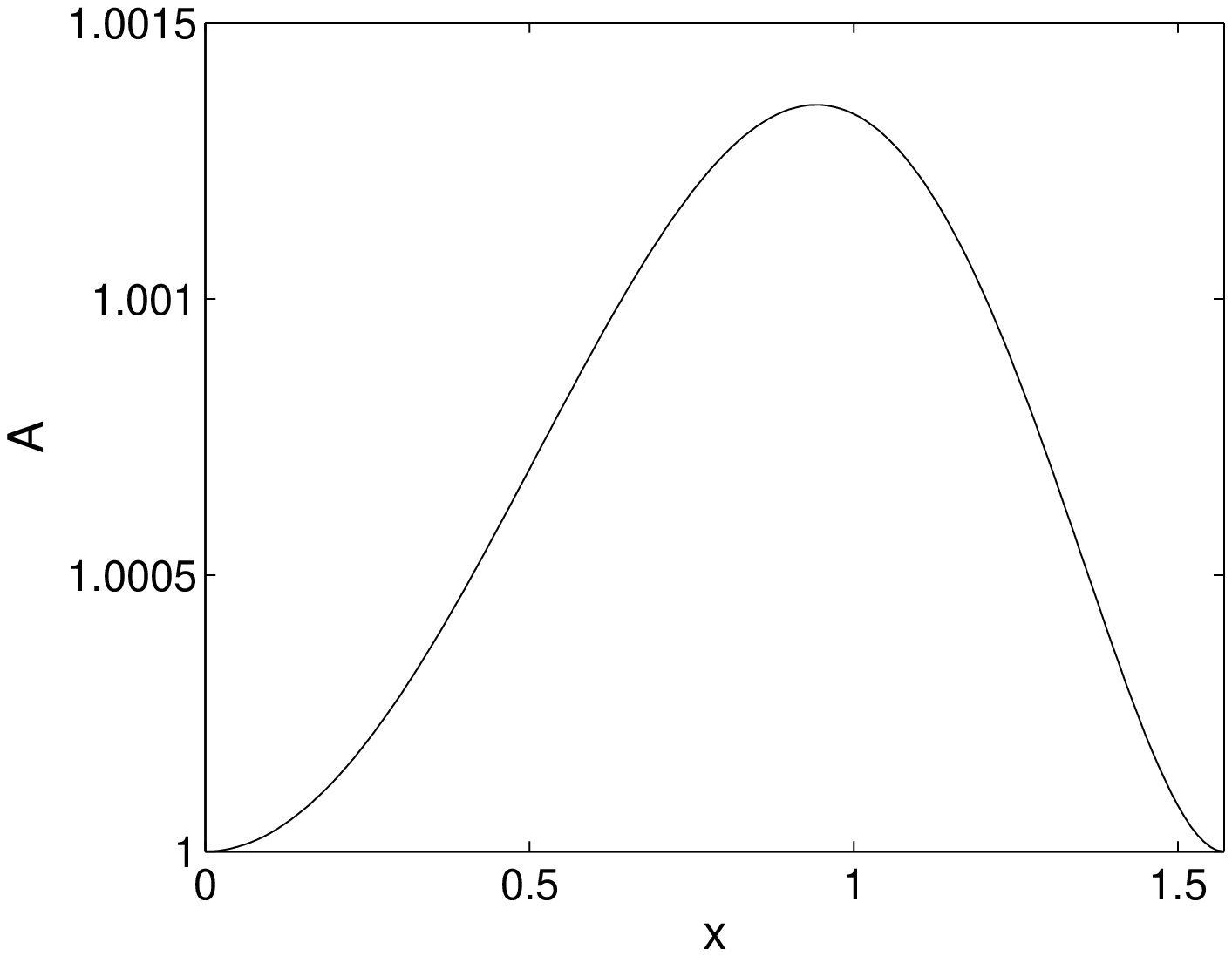}}}
\end{center}
\caption{
Instanton solution with $\phi_0=0.1$.}
\label{instanton}
\end{figure}

\begin{figure}[ht]
\begin{center}
$\begin{array}{c@{\hspace{.2in}}c@{\hspace{.2in}}c} \epsfxsize=2.5in
\epsffile{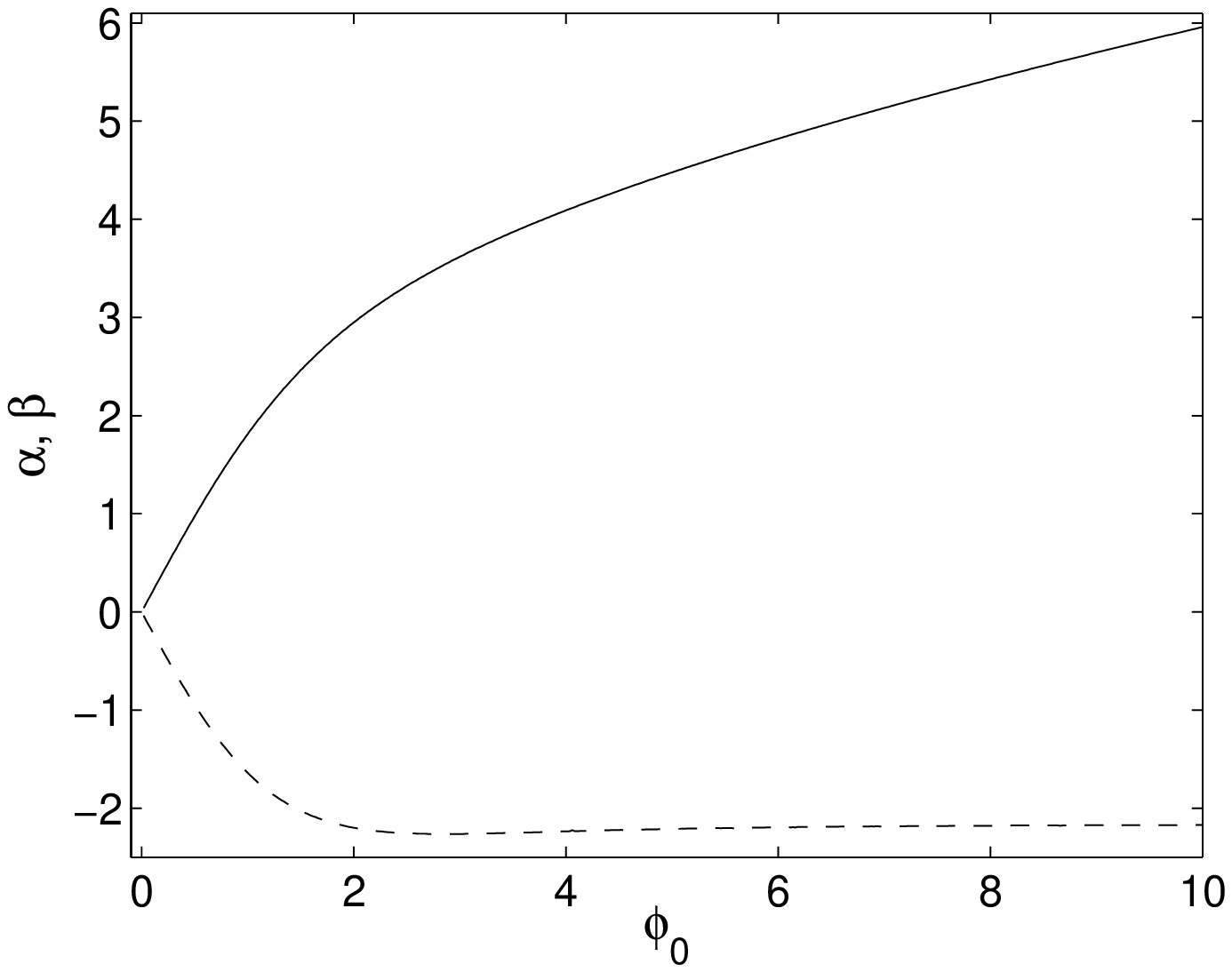}  &
     \epsfxsize=2.5in
    \epsffile{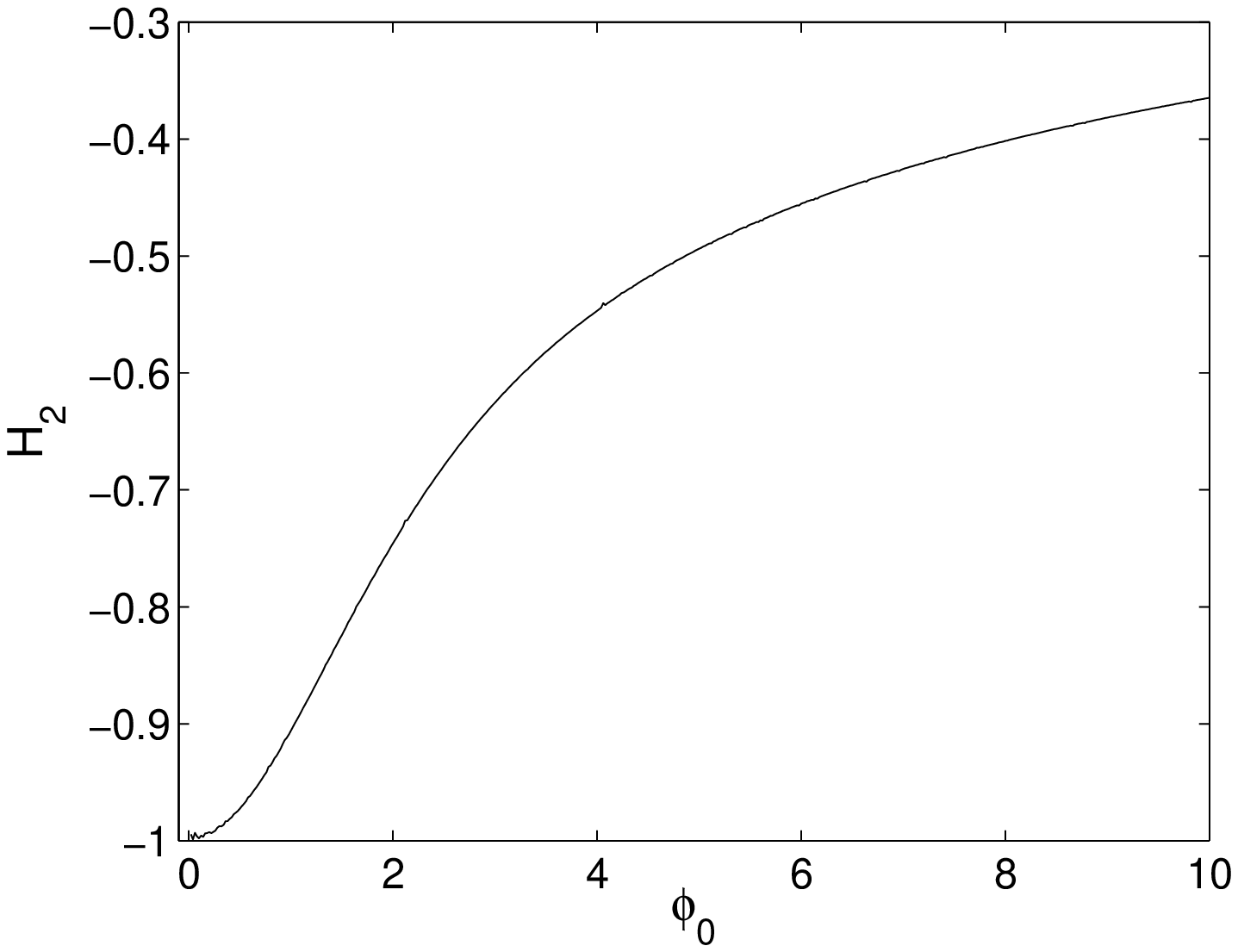}
\end{array}$
\end{center}
\caption{
Left panel: $\alpha$ (solid line) and $\beta$ (dashed line) for the
instanton as function of $\phi_0$. 
Right panel: $H_2=\beta/\alpha$ as function of $\phi_0$.}
\label{instanton2}
\end{figure}

Using the $\rho$ coordinate, the equation of motion for the instanton 
is
\beq
b^2 = \frac{2 \rho^2 V(\phi) -6}{\rho^2  (\phi')^2 -6} \, , 
\qquad
b^2  \phi''  +  \left( b b'  +\frac{3 b^2 }{\rho} \right) \phi' - V'(\phi) =0 \, .
\eeq
For numerical convenience we make a change of variables as follows;
instead of $\rho$ we will use $x=\tan^{-1} \rho$; and instead of $b$,
$A=b^2/(\rho^2+1)$. 
The equation of motion for the instanton then reads
\begin{equation}
\phi'' +\left( \frac{2+\cos^2 x}{\sin x\cos x} +\frac{A'}{2 A} \right) \phi' 
-\frac{V'(\phi)}{A \, \cos^2 x} = 0 \, ,
\qquad
A=\frac{2 V(\phi) \sin^2 x  -6 \cos^2 x}{\sin^2 x\cos^2 x (\phi')^2 -6} \, .
\label{eq:inst1}
\end{equation}
which is sometimes conveniently written as
\begin{equation}
\phi'' = \frac{1}{A} \left(\frac{V'}{\cos^2 x} +\frac{\sin 2x}{6} (V+3) \phi'  \right)-
 \frac{2+\cos^2 x}{\sin x\cos x} \phi'
+\frac{\sin2x}{12} \left(2+\sin^2 x\right) (\phi')^3 \, .
\label{eq:inst2}
\end{equation}
Fig.~\ref{instanton} shows an example of an instanton
solution. The family of solutions is parametrized by the
value $\phi_0\equiv\phi(0)$, i.e.~the value of the scalar field at the
origin.  
Each of these solutions corresponds to a particular value of
$(\alpha,\beta)$ of eq.~\eqref{asintotico} at infinity; see
fig.~\ref{instanton2}. 
The limit $\phi_0 \rightarrow 0$ corresponds to $H_3 \rightarrow \pm
\infty$ for tripletrace deformations, while for doubletrace
deformations $H_2\rightarrow -1$. 

The decay rate can be computed in terms of the Euclidean action of the 
instanton, using the Coleman-de Luccia \cite{CdL} method. The relevant
quantity determining the tunneling rate is $e^{-(S-S_{\rm AdS})/\hbar}$,
with $S$ being the instanton action and $S_{\rm AdS}$ the void AdS
action.  
In the tripletrace case, it was checked in \cite{HH1} that $S-S_{\rm AdS}$
is a finite quantity and hence the tunneling occurs in a finite
time. We checked that this is also the case for the doubletrace
deformation with negative $H_2$.

After the tunneling occurs, the initial condition is obtained by
restricting the instanton solution to the equator.
In the case of AdS-invariant boundary conditions with a tripletrace
deformation, time evolution can be obtained directly via analytical
continuation of the instanton. Inside the lightcone the metric has the
form of a Friedmann-Robertson-Walker solution
\beq
ds^2= - d t^2+ a^2(t) \left( dr^2 + \sinh^2 r \, d \Omega_2^2 \right) \, ,
\eeq
with equations of motion 
\beq
\dot{a}^2=1+\frac{a^2}{3} \left( \frac{\dot{\phi}^2}{2} + V(\phi)  \right)\, ,
\qquad
\ddot{\phi}+ 3 \frac{\dot{a}}{a} \dot{\phi} + V'(\phi) = 0 \, ,
\eeq
describing a universe which is initially expanding but then eventually
contracting; thus developing a big crunch singularity within a finite
time \cite{CdL}.

\subsection{Black holes}

We recall that the Einstein-scalar theory (\ref{azio})
has also  other time-independent solutions than empty AdS.
For all values of $H_2$ and $H_3$ the AdS-Schwarzschild BH is a
solution with the metric 
\beq
ds^2=-\frac{A e^{-2\delta}}{\cos^2 x} \, dt^2 +\frac{d x^2}{A \, \cos^2 x} + \tan^2 x
\, d \Omega^2_2 \, , \qquad 
A = 1 -\frac{M_{\rm BH}}{4 \pi} \frac{\cos^2 x}{\tan x} \, ,
\eeq
with $\delta=0$.
The mass and the temperature as a function of the horizon $x_h$ are
\beq
M_{\rm BH}=4 \pi \frac{\tan x_h}{\cos^2 x_h} \, , \qquad
T_{\rm BH}=\frac{1}{4 \pi} \left( 3 \tan x_h + \frac{1}{\tan x_h }\right) \, ,
\eeq
and the specific heat is negative for $x_h < \pi/6$.

More general BH solutions with scalar hair have been studied
in \cite{Hertog:2004dr}; among the time-independent solutions, there
are also solitons without horizons, see e.g.~\cite{HH1,Hertog:2004ns}. 
The time-independent equations of motion read
\begin{align}
 \frac{1}{\tan^2 x} \left(\tan^2 x A e^{-\delta} \Psi\right)'- 
 \frac{d \, V}{d \, \phi} e^{-\delta} \frac{1}{\cos^2 x} =0\, ,  
\qquad
\delta'=- \frac{\sin x \cos x}{2} \,\Psi^2 \, , \\
A' = \frac{1+ 2 \sin^2 x}{\sin x \cos x} (1-A) -
 \frac{ \sin x \cos x }{2} A \Psi^2 -(V(\phi)+3) \tan x  \, ,
\end{align}
where $\Psi\equiv d\phi/dx$. 
At the horizon $x_h$ (defined by  $A(x_h)=0$) the following boundary condition should be
imposed 
\beq
\Psi=\frac{\tan x }{1-V(\phi) \tan^2 x} \frac{1}{\cos^2 x}   \frac{d
  \, V}{d \, \phi}    \, . 
\eeq
All these time-independent solutions \cite{Hertog:2004dr,HH1,Hertog:2004ns}
have a negative ratio $H_2=\beta/\alpha < 0$.
The only time-independent solution compatible with $H_2 \geq 0$ is the
Schwarzschild BH, which is a solution for arbitrary $H_2$.

We will next proceed to uncovering the duals of time-dependent
boundary theories; the BH will be a good candidate to screen the
big-crunch singularity for those cases where we will find that
dynamical stabilization occurs. 

\subsection{Time-dependent equations of motion}

We now turn to matching the gravitational solutions to
time-dependent boundary conditions. This requires in turn, that the
bulk solutions be time dependent. To this end we consider a
spherically symmetric setup, with the metric 
\beq
ds^2=-\frac{e^{-2 \delta(x,t)} A(x,t)}{\cos^2 x} dt^2 +\frac{d
  x^2}{A(x,t) \, \cos^2 x} + \tan^2 x \, d \Omega^2_2 \, . 
\label{eq:bizonmetric}
\eeq
as well as a scalar field $\phi(x,t)$.
The cross term $dx \, dt$ is absent from the metric due to the choice
of gauge.
There is still some freedom in shifting the variable $\delta$ by 
a function of time; this corresponds a time re-parametrization.
In the following we will set $\delta=0$ at $x=\pi/2$, such that $t$
is the global time coordinate in asymptotic AdS space.

Let us introduce the auxiliary variables $\Psi=\phi'$ 
and $\Pi=A^{-1} e^{\delta} \dot{\phi}\,$  (where $\,'\,$  denotes
derivative with respect $x$ and $\,\dot{}\,$ with respect to $t$) for
which the equations of motion read
\begin{align}
(A e^{-\delta} \Pi)' = \dot{\Psi} \, ,
 \qquad \frac{1}{\tan^2 x} \left(\tan^2 x A e^{-\delta} \Psi\right)'- 
 \frac{d \, V}{d \, \phi} e^{-\delta} \frac{1}{\cos^2 x} = \dot{\Pi}
 \, ,   \label{ads4sistemone} \\
A' = \frac{1+ 2 \sin^2 x}{\sin x \cos x} (1-A) -
 \frac{ \sin x \cos x }{2} A (\Psi^2+\Pi^2 ) -(V(\phi)+3) \tan x  \, ,
 \label{Aeq}\\
\delta'=- \frac{\sin x \cos x}{2} (\Pi^2 + \Psi^2) \, , \qquad \phi'=\Psi \, ,
\label{deltaeq}\\
\dot{\phi}=\Pi A e^{-\delta} \, , \qquad
 \dot{A} +  (\sin x \cos x) A^2 e^{-\delta} \Psi  \Pi =0 \, . \label{tempoeq}
\end{align}
The last equation of \eqref{tempoeq} is a consequence of the other
ones.

In order for the solution to be smooth, we must require that
$\Psi|_{x=0}=\Pi'|_{x=0}=A'|_{x=0}=\Pi|_{x\to\pi/2}=A'|_{x\to\pi/2}=0$ and 
$A|_{x=0}=A|_{x\to\pi/2}=1$; both $\Pi|_{x=0}$ and $\Psi|_{x\to\pi/2}$
are non-vanishing functions of time. 
The quantities $(\alpha,\beta)$ are directly related to the values of
$\Psi$ and $\Psi'$ at the boundary
\beq
\Psi(\pi/2)=-\alpha \, , \qquad
\Psi'(\pi/2)=2 \beta \, .
\eeq
Near the boundary, $\phi$ is going to zero linearly in $(\pi/2 - x)$, 
while $\Psi$ is finite.
In the following we will specialize our treatment to time-dependent 
boundary conditions of the form 
\beq
W=\frac{H(t)}{2} \mathcal{O}^2 \, , \qquad 
H(t)=H_2(t)=\frac{\beta(t)}{\alpha(t)} \, ,
\label{accona}
\eeq
corresponding to doubletrace operators.
The boundary conditions at $x \rightarrow \pi/2$ are
\beq
\Psi' = - 2 H \Psi \, , \qquad
\Pi''=-2 \dot{H} \Psi -2 H \Pi' \, .
 \label{newespressione}
\eeq

\subsection{Observables\label{sec:obs}}

In this section we will define some useful observables describing
geometric features of the numerical solutions. 
The first observable we address is the scalar curvature, which by
means of the equations of motion, is 
\beq
R = 4 V(\phi)+ A (\Psi^2-\Pi^2) \cos^2 x  \, .
\eeq

The second observable we need is the total energy for which it proves
convenient to express the metric component $g_{rr}$ in terms of
the function $A(x)$. Setting $A(\pi/2)=1$ and $A'(\pi/2)=0$ we get
\beq
g_{rr}= \frac{1}{r^2} - \left( 1+ \frac{A''(\pi/2)}{ 2} \right) \frac{1}{r^4} +
\frac{A^{(3)}(\pi/2) }{6}  \frac{1}{r^5} + \ldots 
\eeq
Expanding eq.~(\ref{Aeq}) in series around $x=\pi/2$,
it follows that $A''(\pi/2)=\Psi^2(\pi/2)=\alpha^2$.
Using an expression for the total energy given in
\cite{Hertog:2004ns}, we obtain 
\beq
M = 4 \pi \left(\frac{A^{(3)}(\pi/2)}{6}+\alpha \beta +\int_0^\alpha \beta(\tilde{\alpha}) d \tilde{\alpha} \right)=
4 \pi \left(\frac{A^{(3)}(\pi/2)}{6}+ \frac{3}{2} H \alpha^2 \right)
 \, . \label{eq:energy}
\eeq
This quantity is in general not constant, because the time-dependent 
multitrace operator acts as an external source of energy for the 
conformal field theory described by the AdS${}_4$ geometry.

A trapped two-surface $\mathcal{S}$ 
(see e.g.~\cite{book})
is defined as a closed surface
with the property that the expansion scalars
\beq
\theta_l= (g^{ab}+l^a n^b + l^b n^a) \nabla_a l_b \, , \qquad
\theta_n= (g^{ab}+l^a n^b + l^b n^a) \nabla_a n_b \, .
\eeq
in each of the two forward-in-time null directions $l_a, n_b$ (normal
to $\mathcal{S}$) are both negative. 
We use the cross normalization $l_a n^a=-1$.
Then, considering a foliation by spacelike three-surfaces $\Sigma_t$,
a point $q \in \Sigma_t$ is said to be trapped if it lies on a
trapped two-surface $\mathcal{S}$ in $\Sigma_t$. 
The apparent horizon in $\Sigma_t$ is thus constructed as the boundary
of the union of all the trapped points. 
Let us consider the following null direction of ingoing and outgoing 
wave fronts in the $(t,x)$ direction
\beq
l_a =\frac{1}{\sqrt{2 A} e^\delta \cos x} \{ -A,-e^\delta,0,0 \} \, , \qquad
n_a =\frac{1}{\sqrt{2 A} e^\delta \cos x}  \{ -A,e^\delta,0,0 \} \, .
\eeq
An explicit calculation gives
\begin{align}
\theta_{l,n} &= \mp \frac{\sqrt{2A}}{\sin x} \, .
\end{align}
This shows that no apparent horizon forms as long as $A$ remains positive.

The metric \eqref{eq:bizonmetric} is an example of polar coordinates, 
which are in general defined by the condition
$\mathcal{K}^\theta_\theta =\mathcal{K}^\phi_\phi=0$, where
$\mathcal{K}$ is the extrinsic curvature tensor. It is a general
feature of polar coordinates that they can not be used to parametrize
events happening inside apparent horizons (see
e.g.~\cite{lectures,book} for a discussion); instead they become 
singular when an apparent horizon forms. Apparent horizons are then
detected by the condition that the metric \eqref{eq:bizonmetric} is
singular \cite{lectures,book}, which is therefore given by a vanishing 
$A$. 
An alternative foliation and metric Ansatz should be used in order
to penetrate the apparent horizons.

\subsection{Numerical results}

In this section we present numerical solutions of
eqs.~(\ref{ads4sistemone}-\ref{tempoeq}) obtained for different
time-dependent boundary conditions, 
specified by the function $H(t)$ in eq.~(\ref{accona}).
As initial condition we use the instanton obtained by
eqs.~(\ref{eq:inst1},\ref{eq:inst2}) restricted to the equator.  
We imagine that $H(t)$ is constant for $t<0$ and that the system is
in a metastable vacuum. We then assume that a tunneling occurs at
$t=0$ from where we introduce time dependence to the doubletrace
deformations. 
As a parametrization of the instanton solutions, we fix $\phi_0$,
corresponding to the value of the scalar field at the origin of the
instanton. The relation between $\phi_0$ and the initial value 
of $H(t)=H_2$ is shown in fig.~\ref{instanton2}.

\begin{figure}[!htp]
\centering{}
\includegraphics[width=0.6\linewidth]{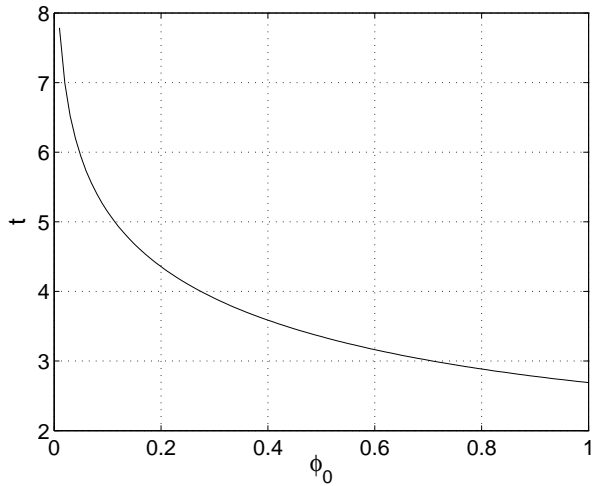}
\caption{The time it takes for an instanton to evolve
 into a big crunch singularity for constant and negative $H(t)=h_0<0$
 as function of $\phi_0$.} 
\label{crunchtime}
\end{figure}

We first consider the case of constant $H(t)=h_0<0$ for which the
solution corresponds to big crunch geometries.
A curvature singularity (which extends to the boundary) develops in
a finite time.
The energy can be computed via eq.~(\ref{eq:energy}) which we have
checked is diverging at the time when the crunch singularity forms. 
A plot of the time it takes for each instanton to evolve into a crunch
as function of $\phi_0$ is shown in fig.~\ref{crunchtime}.
This quantity increases with the steepness of the potential. 
The reason for this is that the larger the value of $\phi_0$ is, the
larger is the energy which causes the crunch.
Note on the other hand the smaller the value of $\phi_0$ is, the
faster is the decay of the vacuum. This behavior can be seen in the 
right-hand part of fig.~\ref{instanton2}.

\begin{figure}[tp]
\begin{center}
\mbox{\subfigure[$\phi(x,t)$]{\epsfxsize=2.8in \epsffile{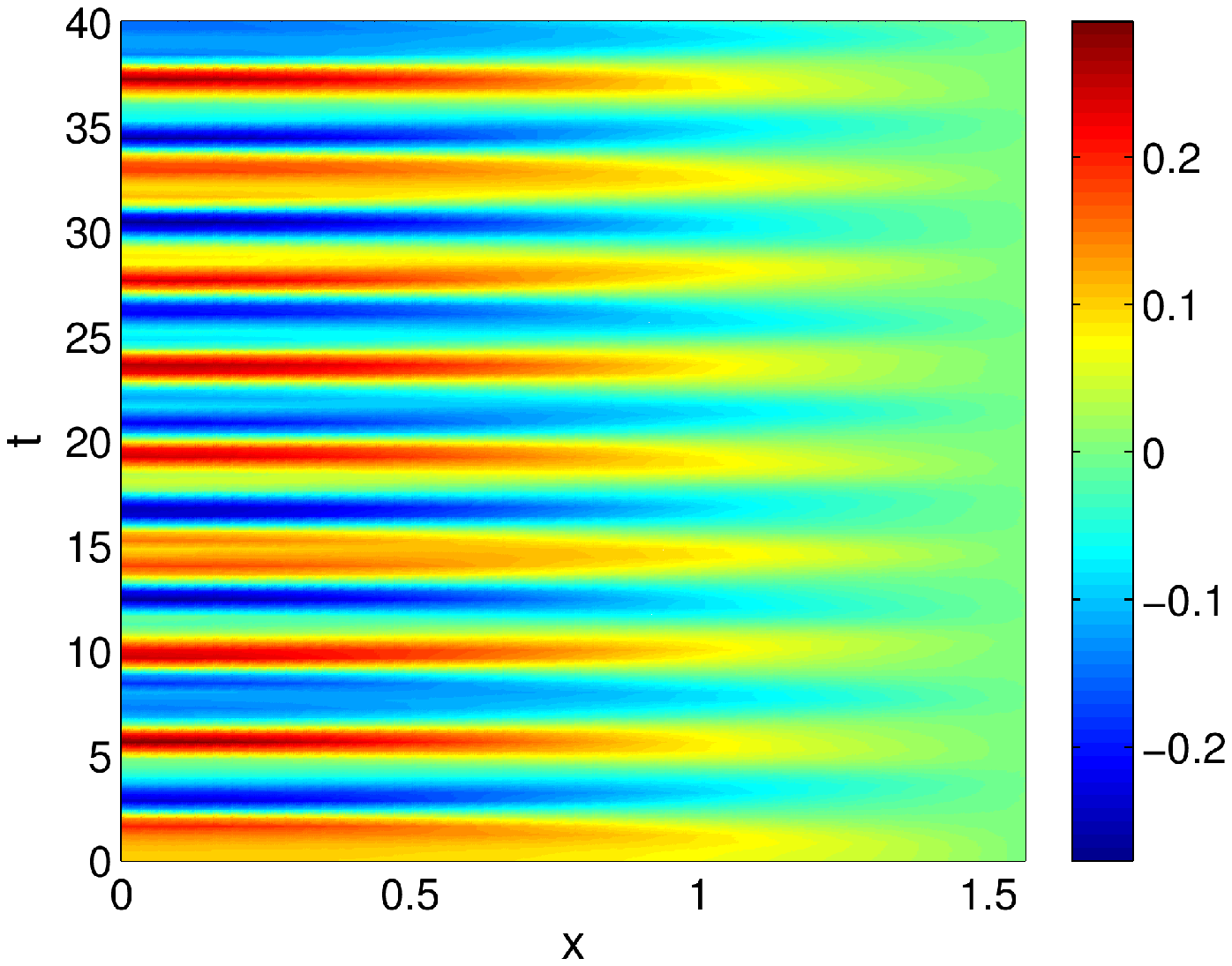}}\quad
\subfigure[$A(x,t)$]{\epsfxsize=2.8in \epsffile{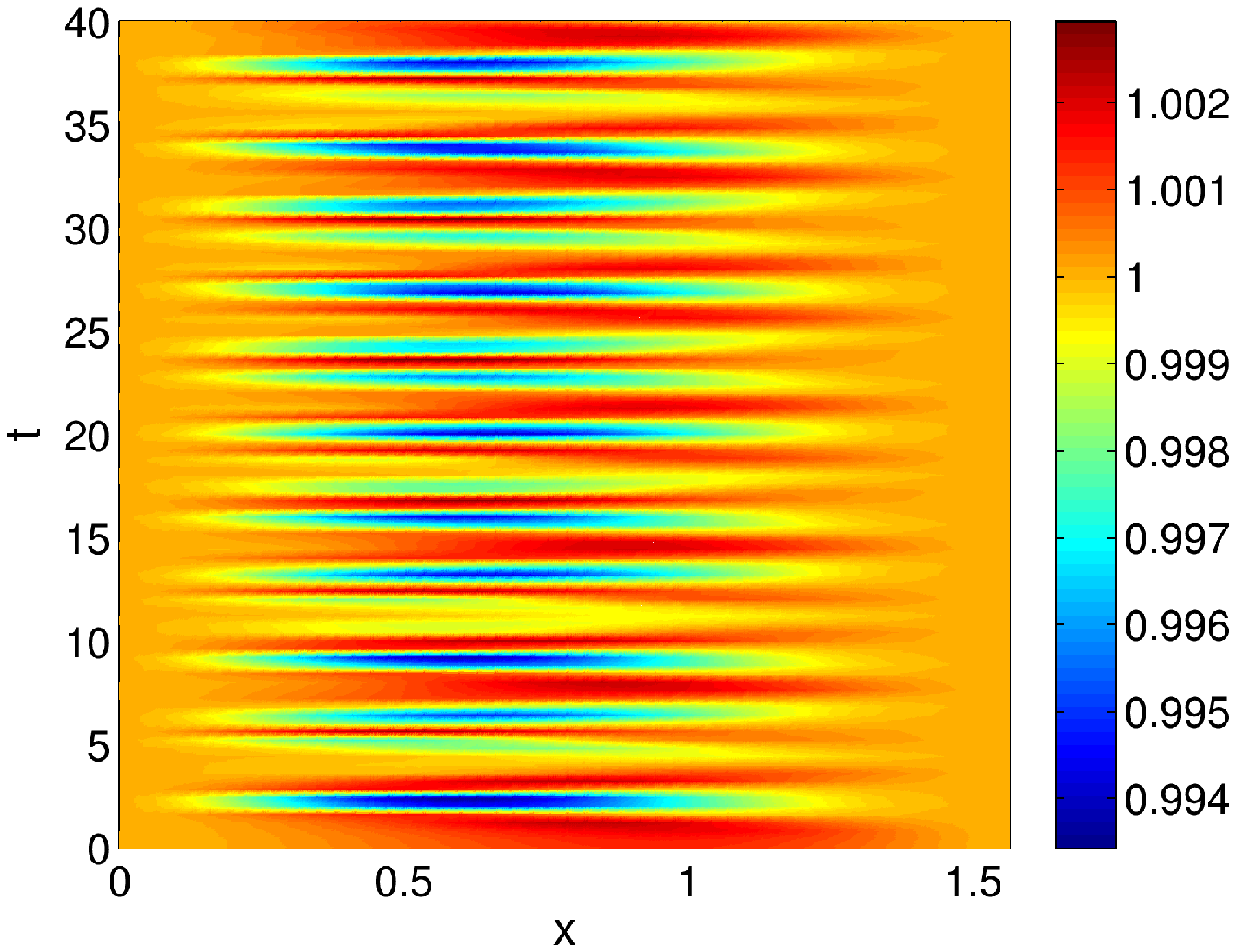}}}
\mbox{\subfigure[$\delta(x,t)$]{\epsfxsize=2.7in \epsffile{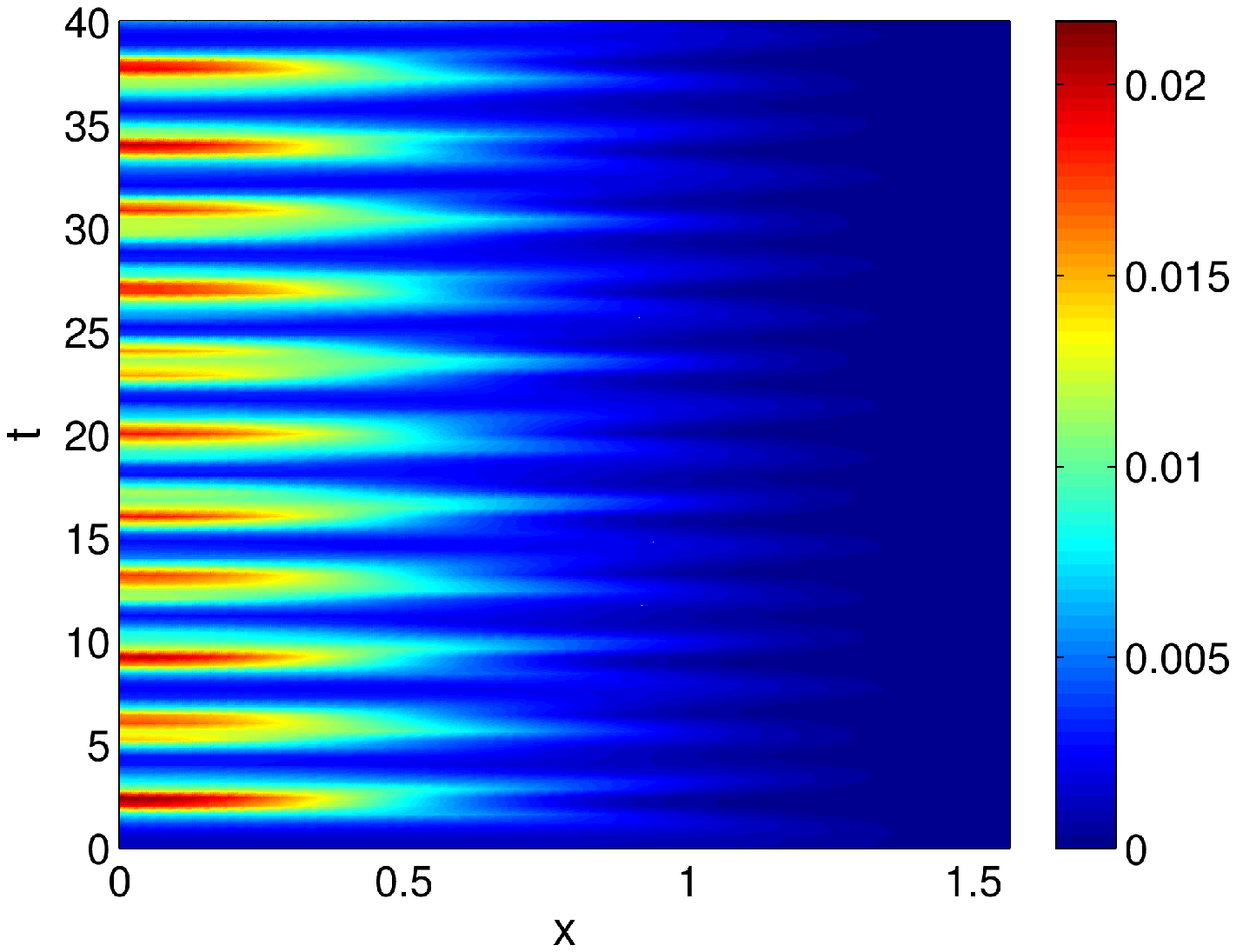}}\quad
\subfigure[scalar curvature $R(x,t)$]{\epsfxsize=2.7in \epsffile{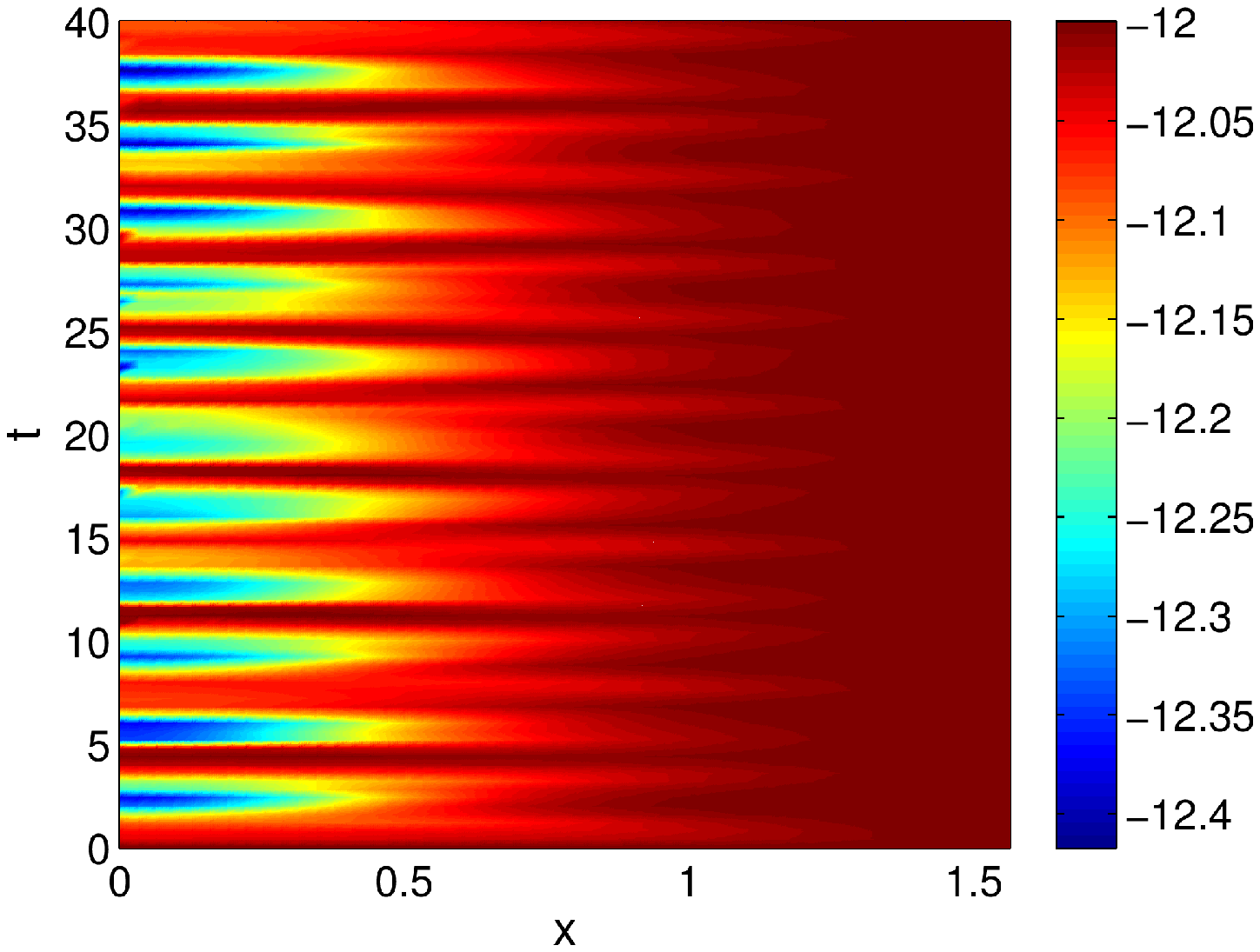}}}
\end{center}
\caption{
A solution with boundary condition \eqref{gradino} and
$\phi_0=0.1$. 
The initial negative unbounded potential is lifted to a constant
positive definite one in a finite time $t\approx 2$ 
much before the time when the crunch singularity would have formed
if the unbounded potential were kept constant (see figure
\ref{crunchtime}). 
The resulting configuration in this case is just a scalar wave 
reflected by the AdS boundary conditions.
}
\label{step1}
\end{figure}

Let us next consider a boundary condition of the type
\beq
H(t) = h_0 \left(2 e^{-t^2} -1\right) \, , \qquad h_0 < 0 \, ,
\label{gradino}
\eeq
which from the boundary point of view corresponds to a negative and
unbounded potential at the initial time, but is then lifted
in a finite time $\approx 2$ to a positive definite one.
A solution within this class is shown in fig.~\ref{step1}. 
For a massless particle in AdS space, it takes the time $\pi$ to reach
the boundary and come back to the origin, corresponding to a
frequency $\omega_0=2$. Indeed in the plots both the fields and the
curvature oscillate with a frequency of roughly $2$. 

\begin{figure}[!htp]
\centering{}
\includegraphics[width=0.6\linewidth]{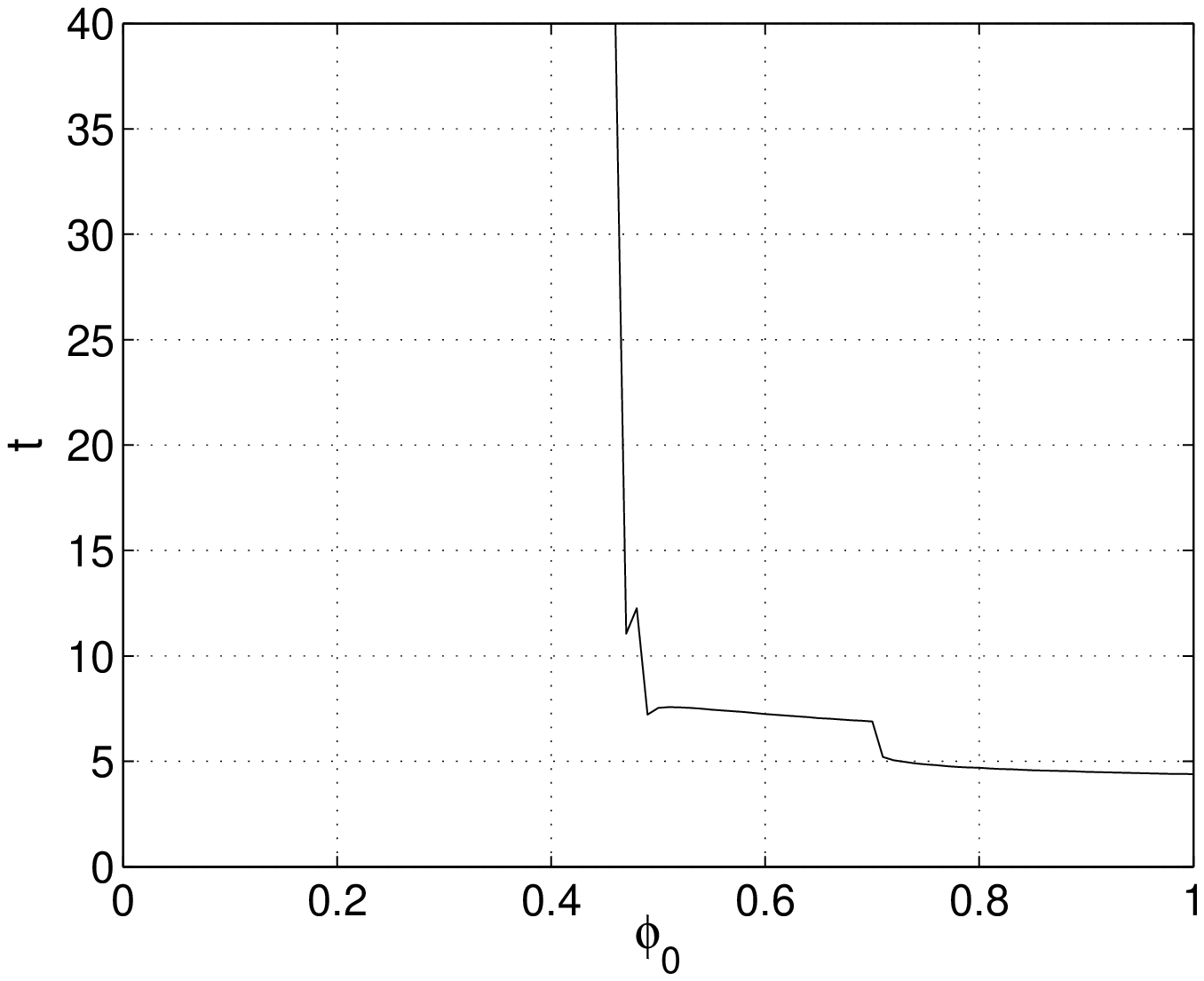}
\caption{The time after which a BH is
  formed as function of $\phi_0$, for the boundary condition
  (\ref{gradino}). The simulation is run until $t_{\rm max}=40$.
  For $\phi_0 \leq 0.46$ we notice a very sharp transition:
  no horizon forms before $t_{\rm max}$;
  for $\phi_0=0.46$, it is found that
  a BH horizon forms at $t\approx 882$. } 
\label{BHtime}
\end{figure}

At late times we expect a BH to form due to the weak turbulent
instability discussed in \cite{Bizon,Dias,deOliveira:2012ac}.  
However, in the regime of small $\phi_0$ the horizon forms only
after a very long time and after the scalar field
has bounced off the AdS boundary many times.
The time of BH formation as function of $\phi_0$ is shown in
fig.~\ref{BHtime}. 
For computational constraints we computed the solutions only till
time $t_{\rm max}=40$.
In the cases where a horizon has formed during the finite calculation
time, we have checked that the final configuration is, to a good  
approximation, a Schwarzschild BH outside the horizon.

\begin{figure}[!htp]
\centering{}
\includegraphics[width=0.6\linewidth]{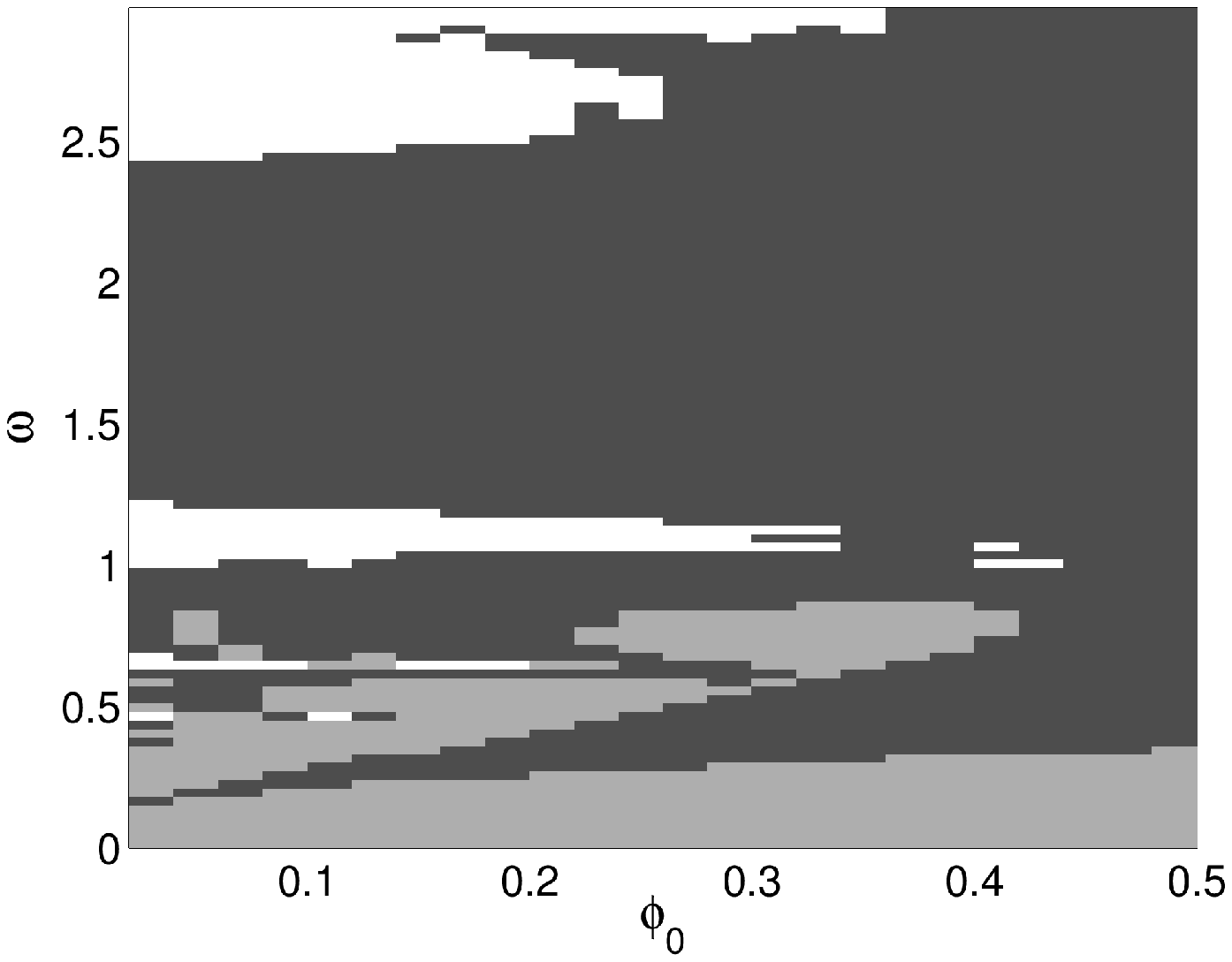}
\caption{Shown in light gray, points in parameter space giving rise 
  to a crunch before $t_{\rm max}=40$, for the boundary condition in
  eq.~(\ref{perizo}). 
  The points for which a BH is formed before $t_{\rm max}$ are
  shown in dark gray. 
  Points in parameter space for which neither a crunch nor a BH is
  formed before $t_{\rm max}$ are shown in white. } 
\label{crunchoso}
\end{figure}

We are now ready to study the boundary conditions related to dynamical
stabilization. These alternate between a negative, unbounded potential
and a positive one  
\beq
H(t) = h_0 \cos(\omega t) \, , \qquad h_0 < 0 \, .
\label{perizo}
\eeq
For sufficiently small $\omega$ a crunch will occur, otherwise given a
general point in the parameter space $(\phi_0,\omega)$ a crunch may or
may not form at some point in the future.
A plot of the parameters for which a big crunch singularity forms
(before time $t_{\rm max}=40$) is shown in fig.~\ref{crunchoso}.

The transition between the crunch and the BH regime
should correspond to a BH with infinite mass.
This regime however is  challenging to study numerically as
it is sensitive to the UV cutoff $x_{\rm max}$ which in turn requires
increasingly small finite difference steps.

\begin{figure}[ht]
\begin{center}
\mbox{\subfigure[$\phi(x,t)$]{\epsfxsize=2.7in \epsffile{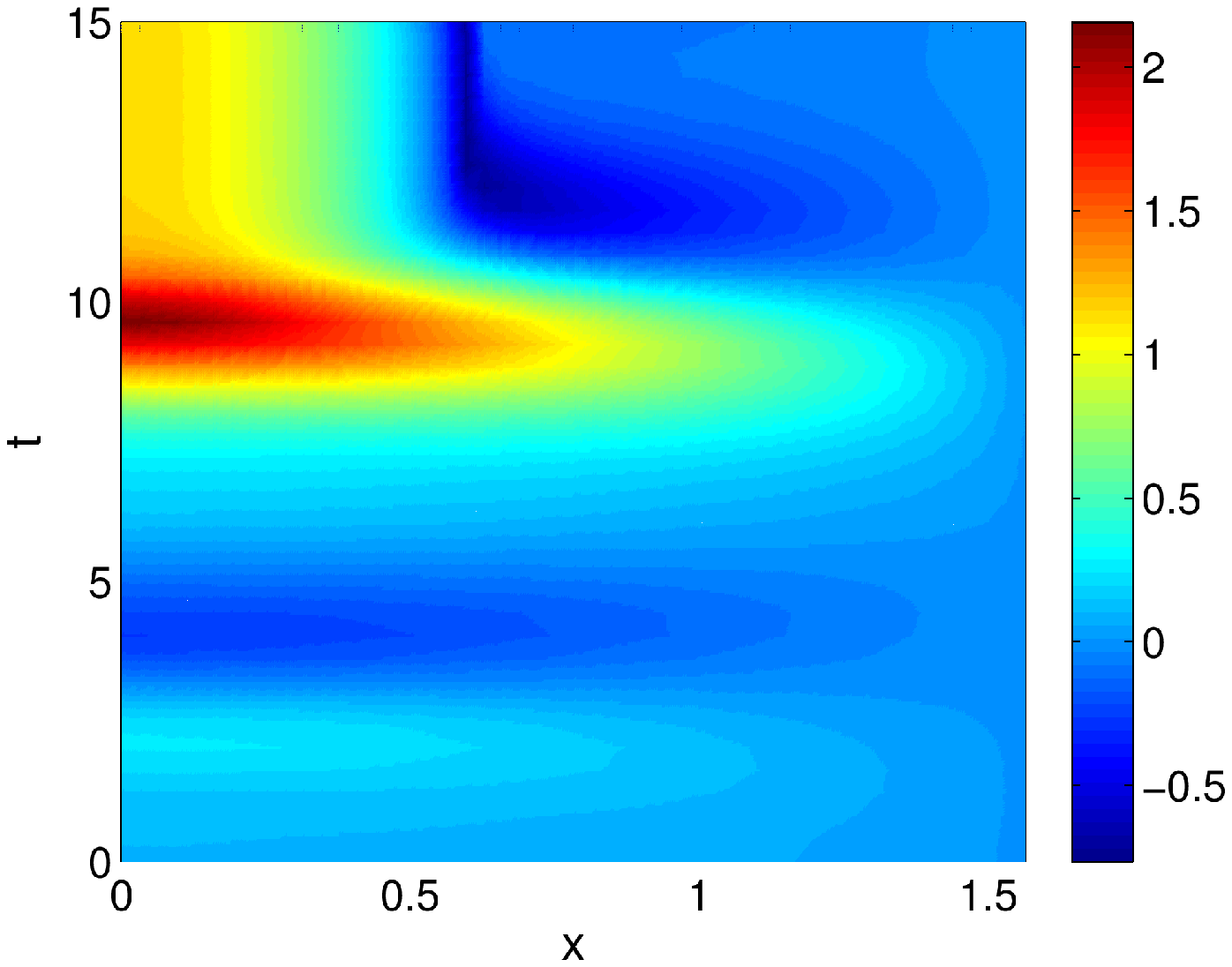}}\quad
\subfigure[$A(x,t)$]{\epsfxsize=2.7in \epsffile{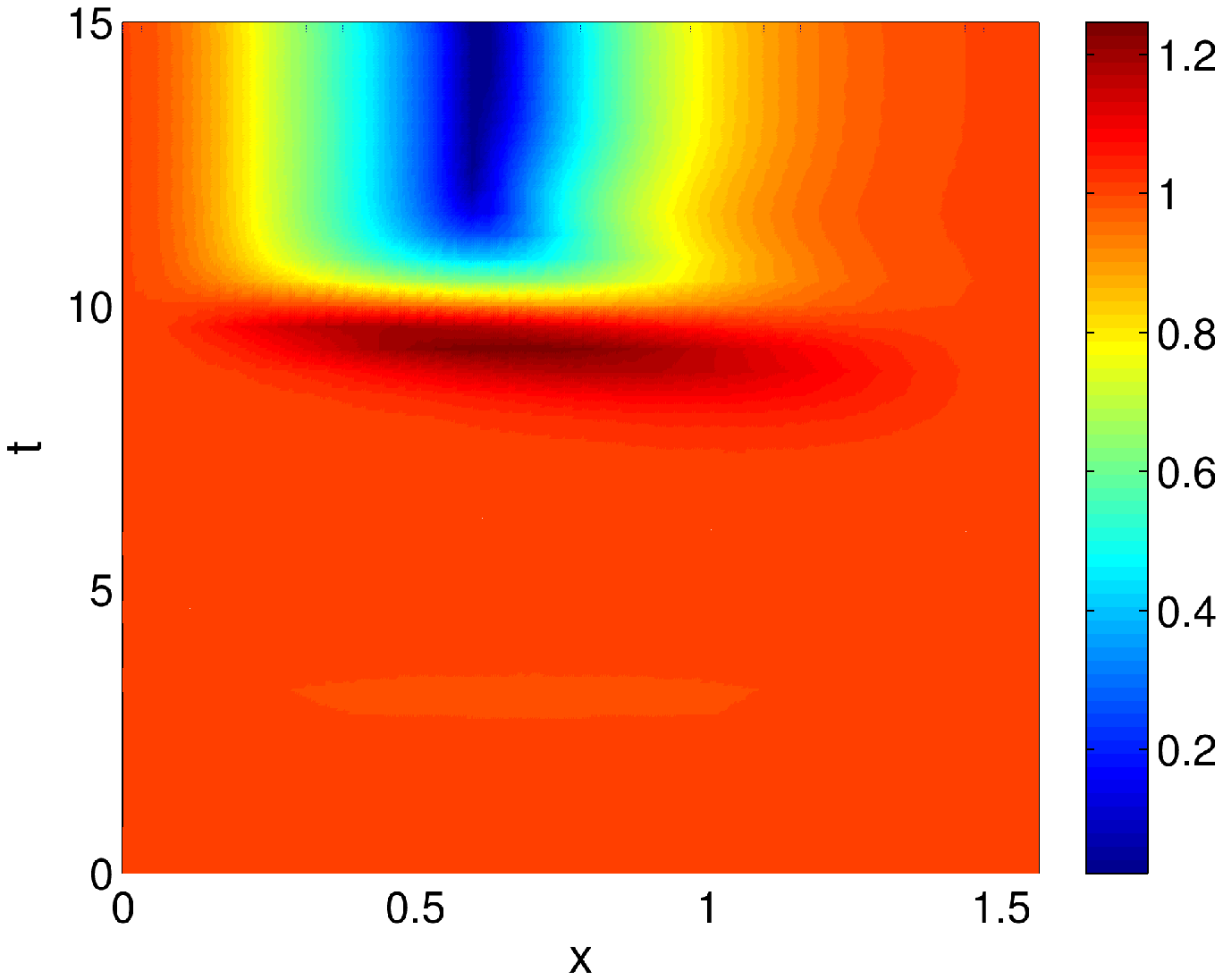}}}
\mbox{\subfigure[$\delta(x,t)$]{\epsfxsize=2.7in \epsffile{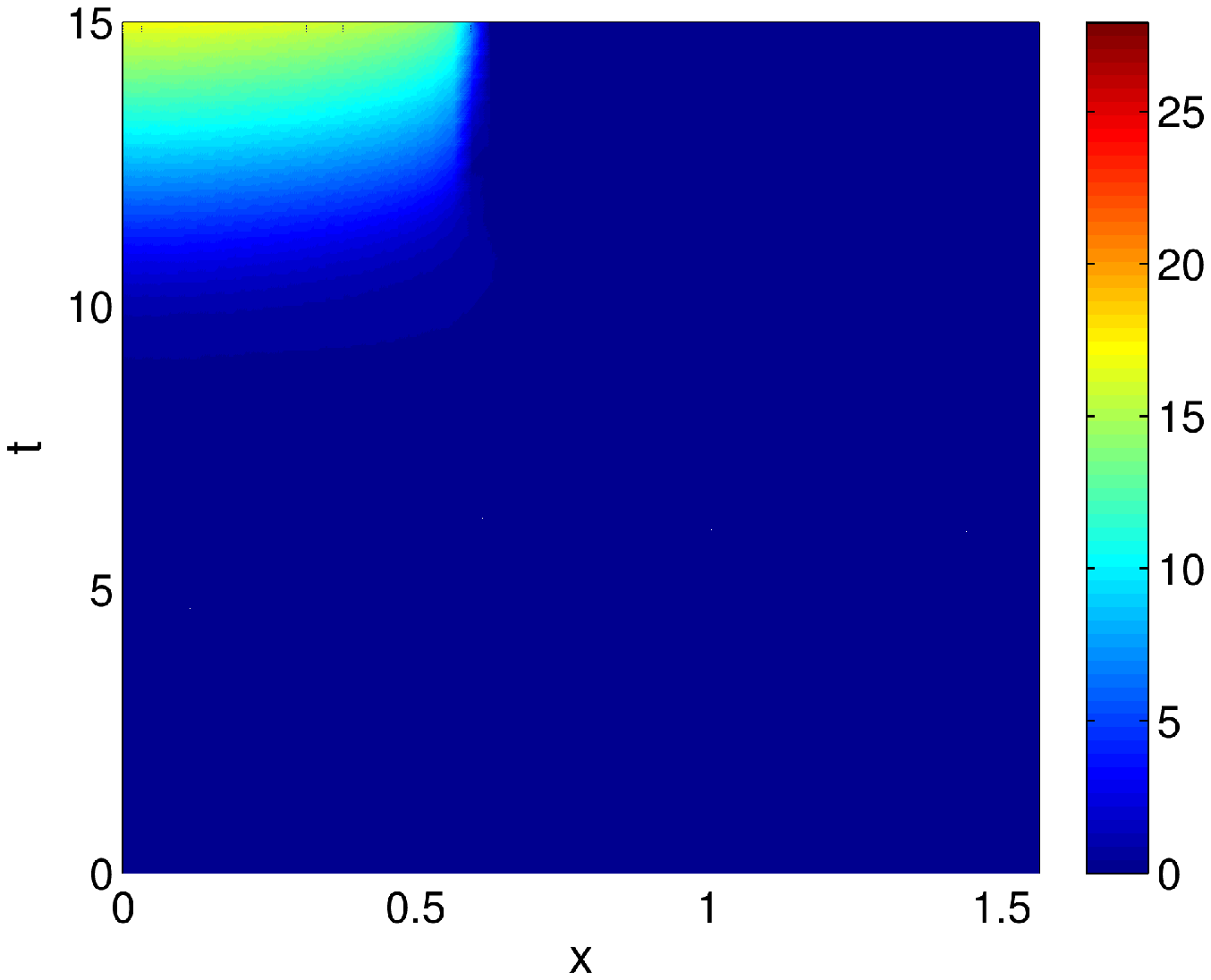}}
\subfigure[scalar curvature $R(x,t)$]{\epsfxsize=2.7in \epsffile{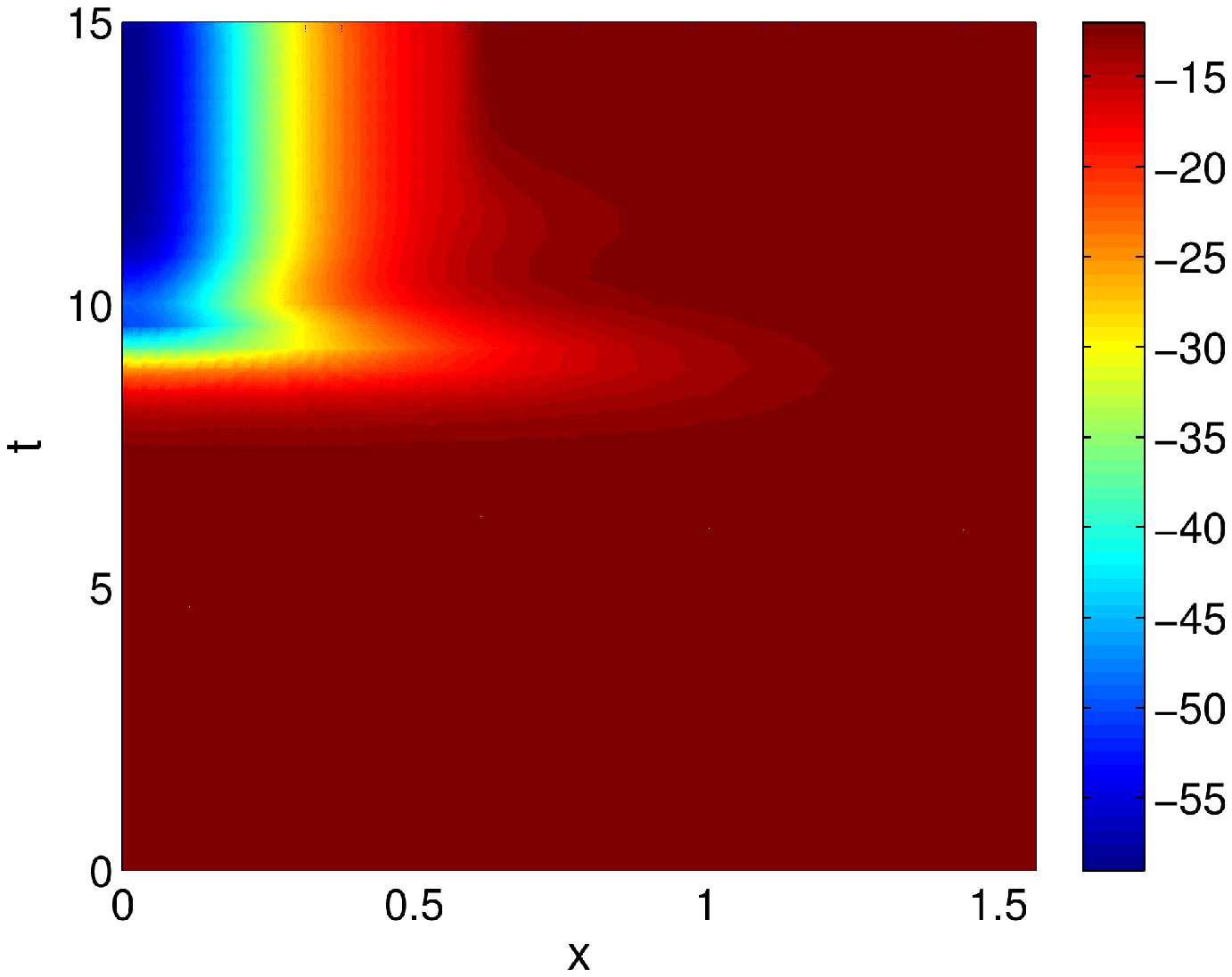}}}
\end{center}
\caption{
A solution where a BH forms for $\phi_0=0.1$
and $\omega=0.9$, corresponding to a dark gray point in
fig.~\ref{crunchoso}. 
Dynamical stabilization is at work:
the big crunch singularity does not reach the boundary in a finite
time as it is screened by an apparent horizon, which is where $A=0$. 
The absolute value of the scalar curvature (d) becomes large
(signaling the formation of a singularity), but it happens inside the 
horizon. }
\label{omega-0e8}
\end{figure}

An example of a solution where a horizon forms is shown in
fig.~\ref{omega-0e8}. As explained in section \ref{sec:obs}, a horizon
can be detected by virtue of a vanishing $A$. 
The scalar curvature $R$ becomes quite large inside the horizon. 
The time evolution after the time where $A$ vanishes cannot be
trusted, because the metric becomes singular and hence cannot be used
to parametrize events inside the apparent horizon. 

\begin{figure}[ht]
\begin{center}
\mbox{\subfigure[$\phi(x,t)$]{\epsfxsize=2.7in \epsffile{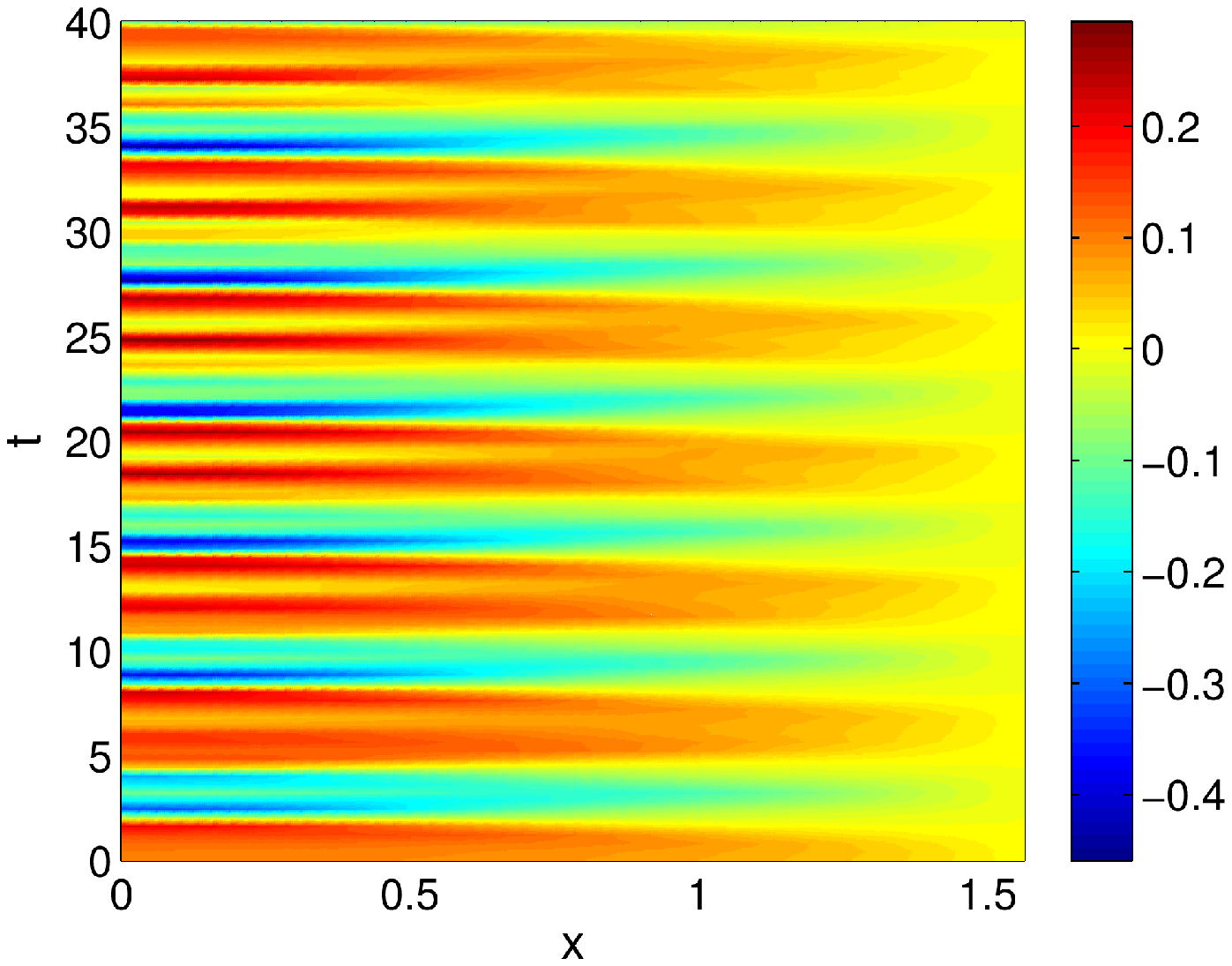}}\quad
\subfigure[$A(x,t)$]{\epsfxsize=2.7in \epsffile{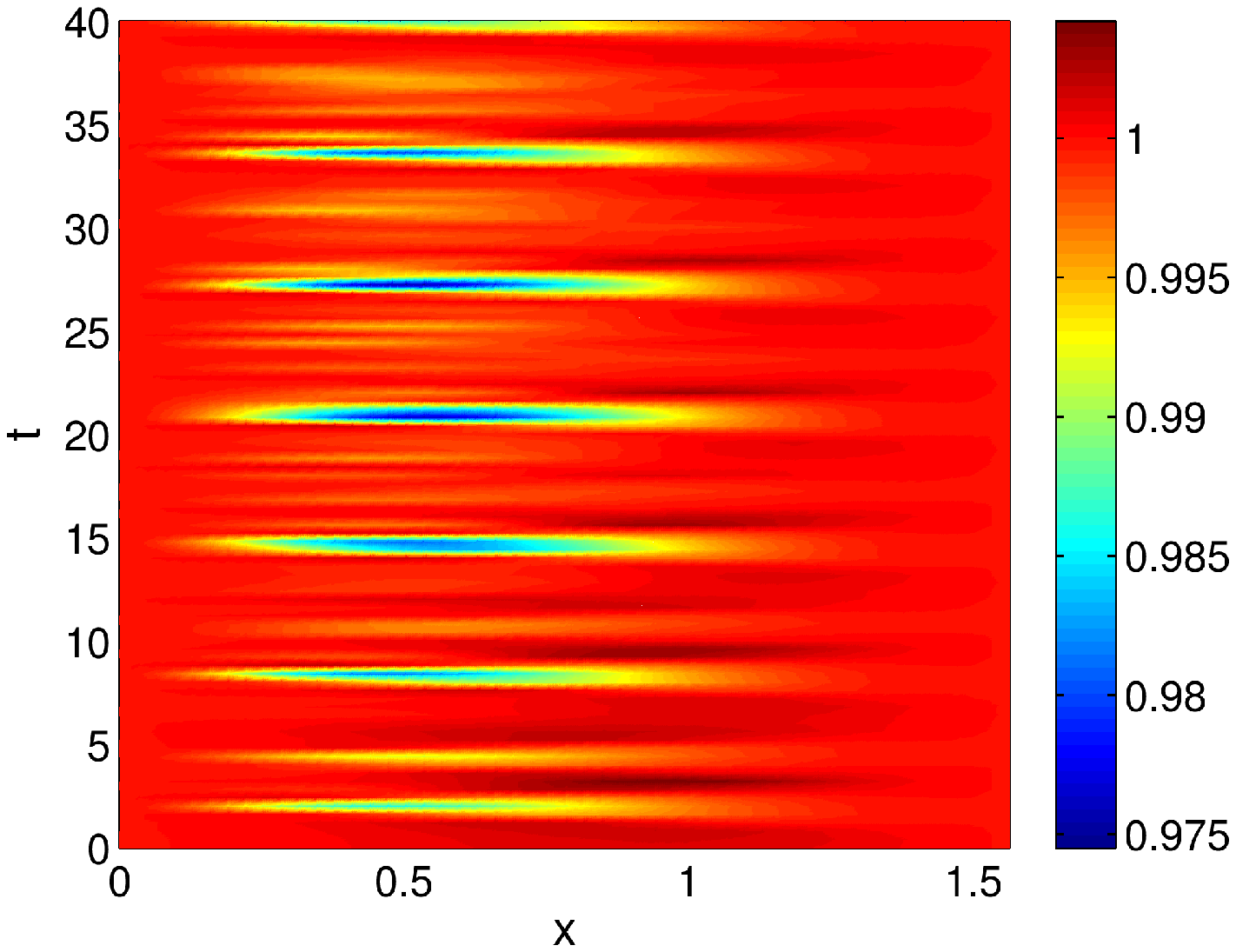}}}
\mbox{\subfigure[$\delta(x,t)$]{\epsfxsize=2.7in \epsffile{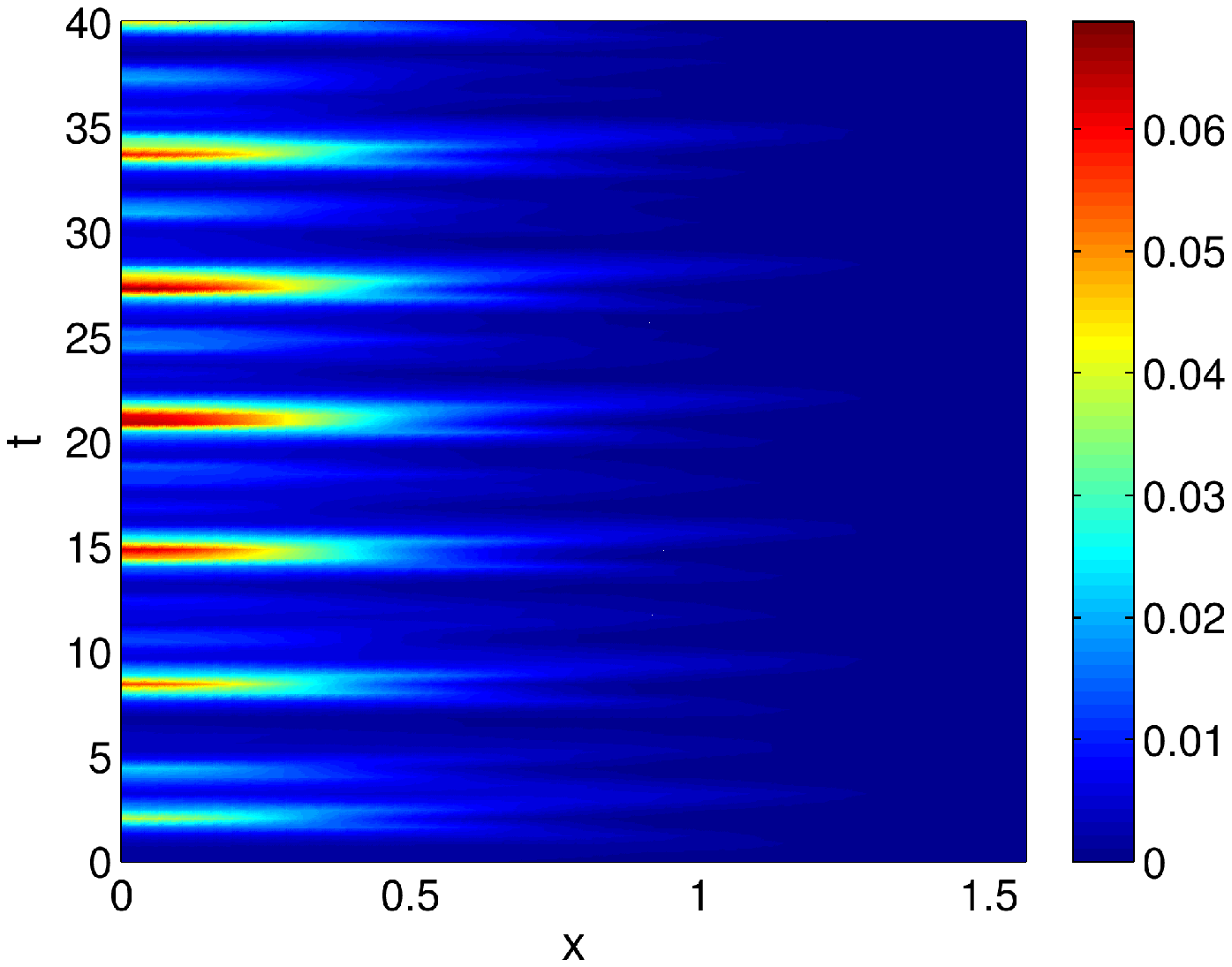}}\quad
\subfigure[scalar curvature $R(x,t)$]{\epsfxsize=2.7in \epsffile{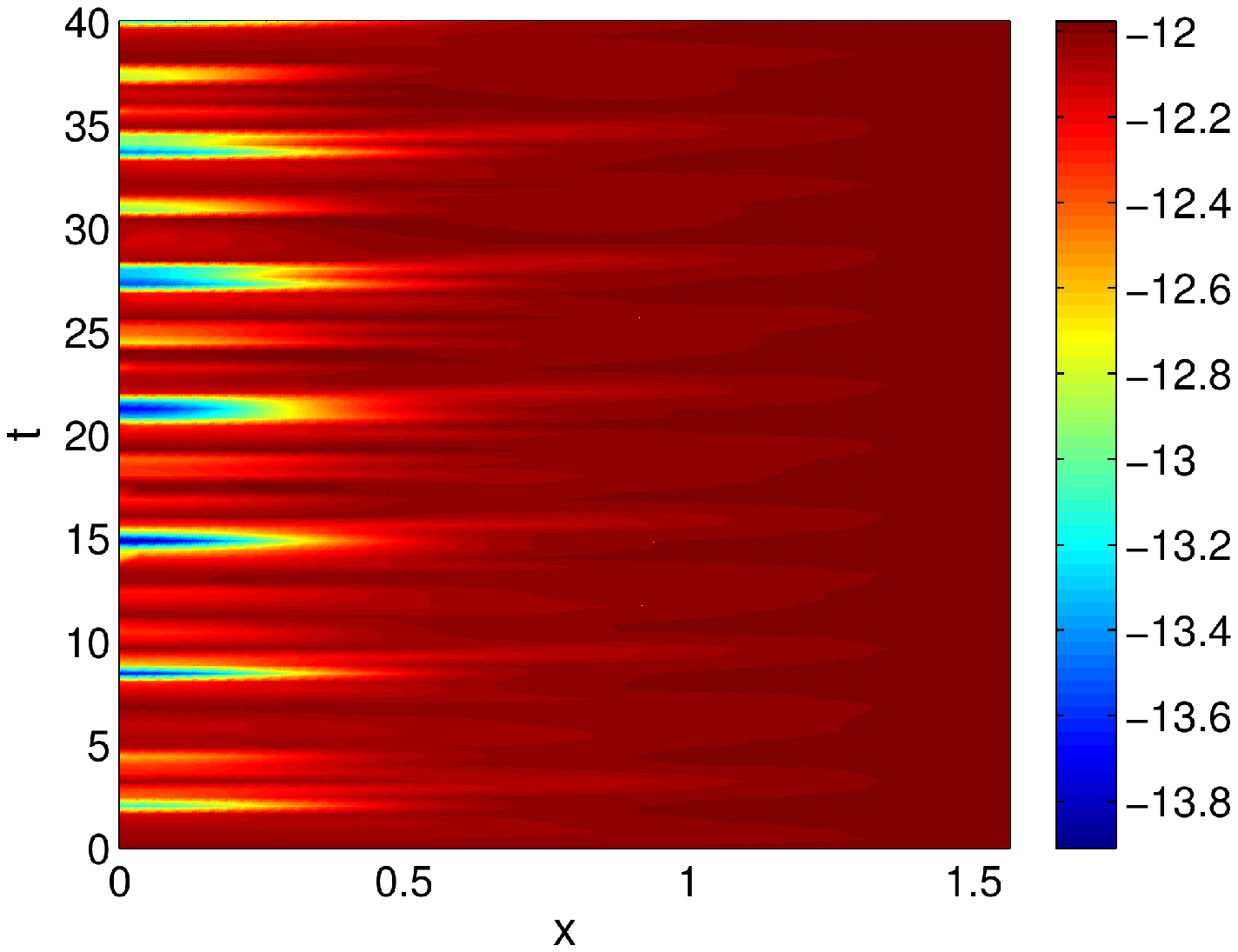}}}
\end{center}
\caption{
A solution where no BH forms
before $t_{\rm max}=40$ for $\phi_0=0.1$ and $\omega=3.0$,
corresponding to a white point in fig.~\ref{crunchoso}. 
Dynamical stabilization is also at work here,
even if no BH horizon has formed before $t_{\rm max}$.  The solution
corresponds to a scalar wave being reflected many times by the
time-dependent boundary condition, without a dramatic increase of the
total energy. The field configuration is similar to that of
fig.~\ref{step1}, even though the stabilization in the latter case 
is not dynamical.}
\label{omega-3e0}
\end{figure}

\begin{figure}[!htp]
\centering{}
\includegraphics[width=0.6\linewidth]{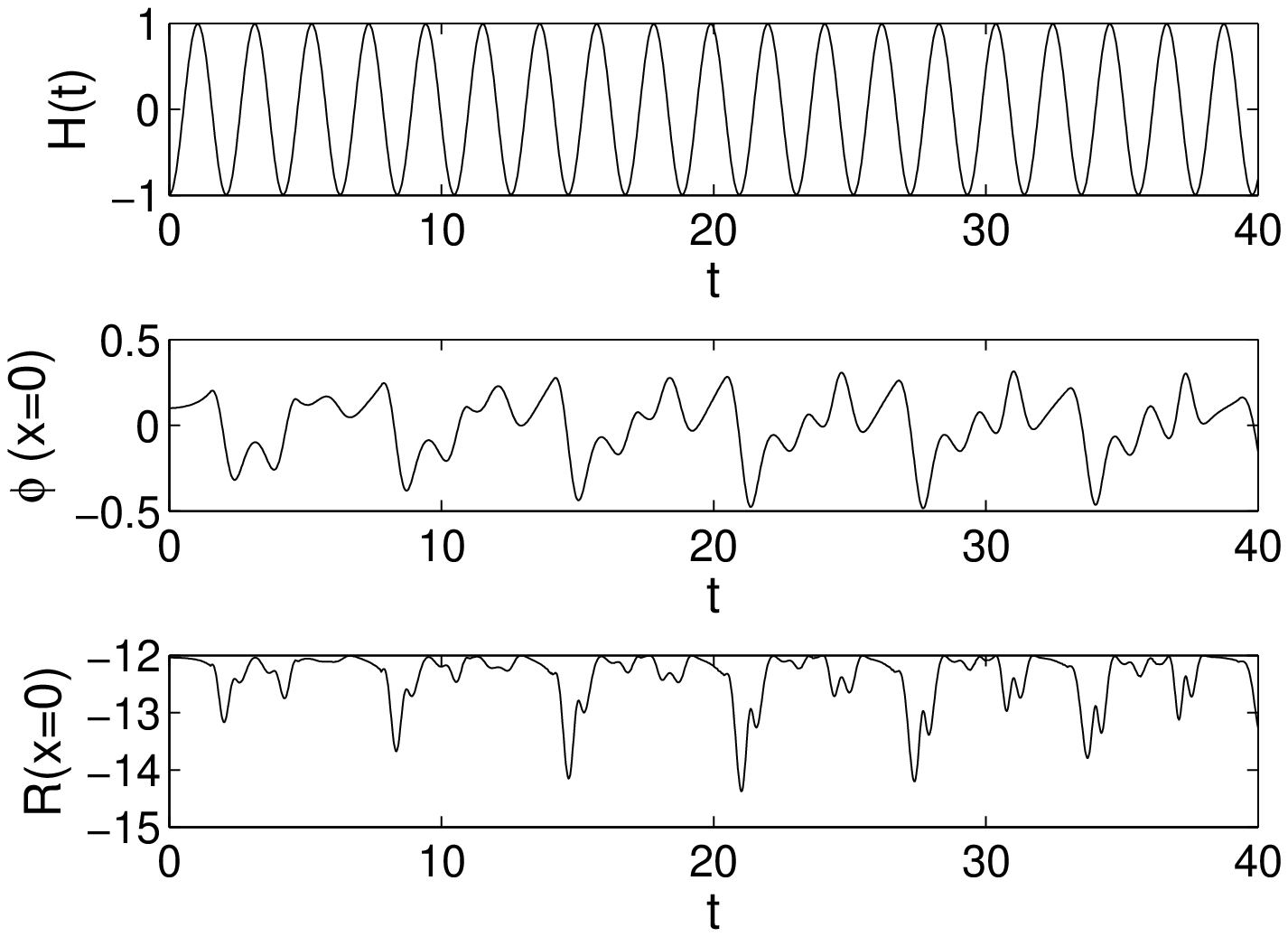}
\caption{The value of $\phi$ and the scalar curvature $R$ at $x=0$ 
are shown as functions of time for $\omega=3$ and $\phi_0=0.1$.
For comparison also the driving potential $H(t)$ is shown.}
\label{oscill}
\end{figure}

An example of a parameter point where no horizon forms before time
$t_{\rm max}=40$ is shown in fig.~\ref{omega-3e0}. Here the scalar
curvature remains rather small 
everywhere. The curvature and the scalar field at $x=0$ are shown as
function of time in fig.~\ref{oscill}.

\begin{figure}[ht]
\begin{center}
$\begin{array}{c@{\hspace{.2in}}c}
\epsfxsize=2.7in \epsffile{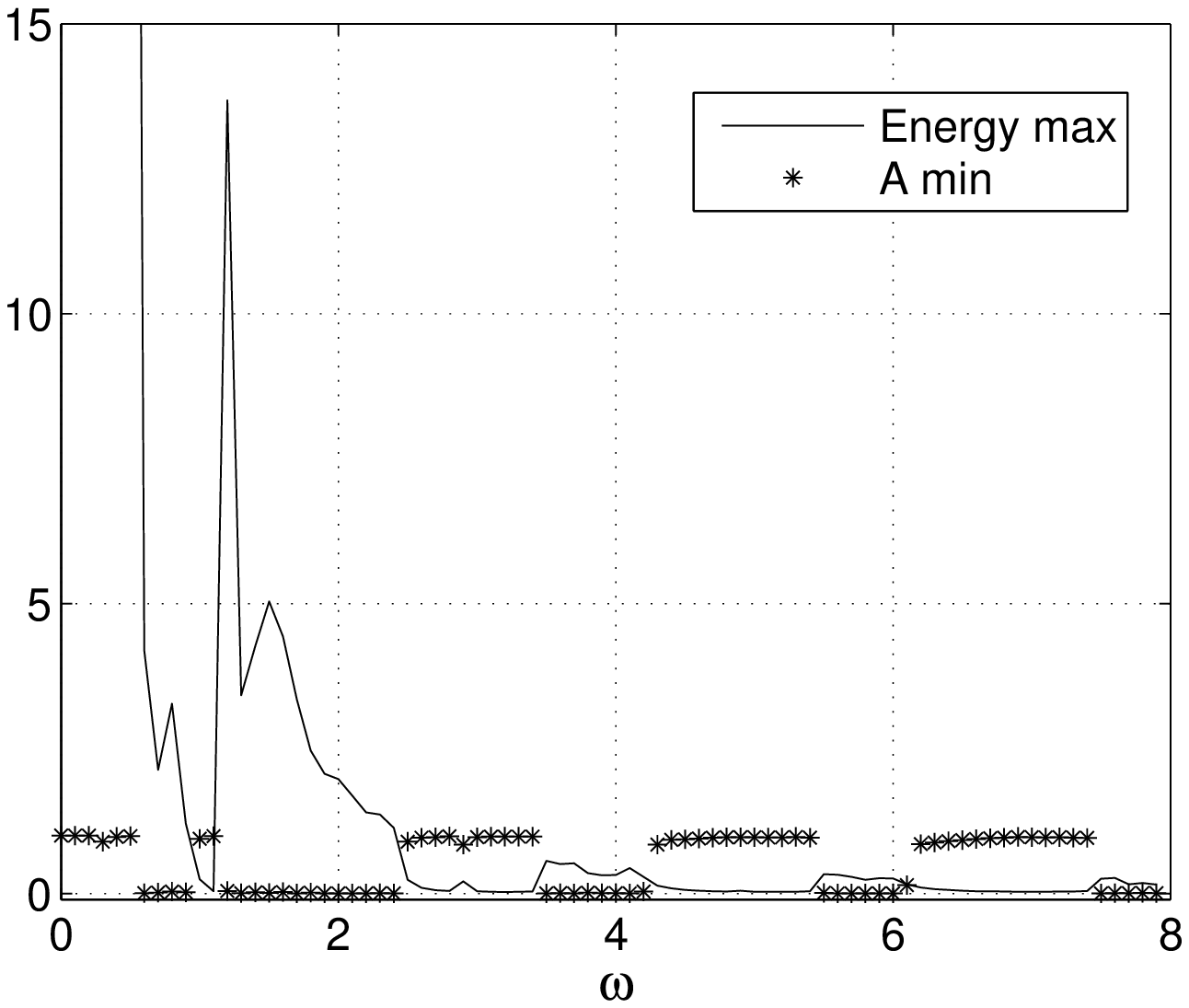}&
\epsfxsize=2.7in \epsffile{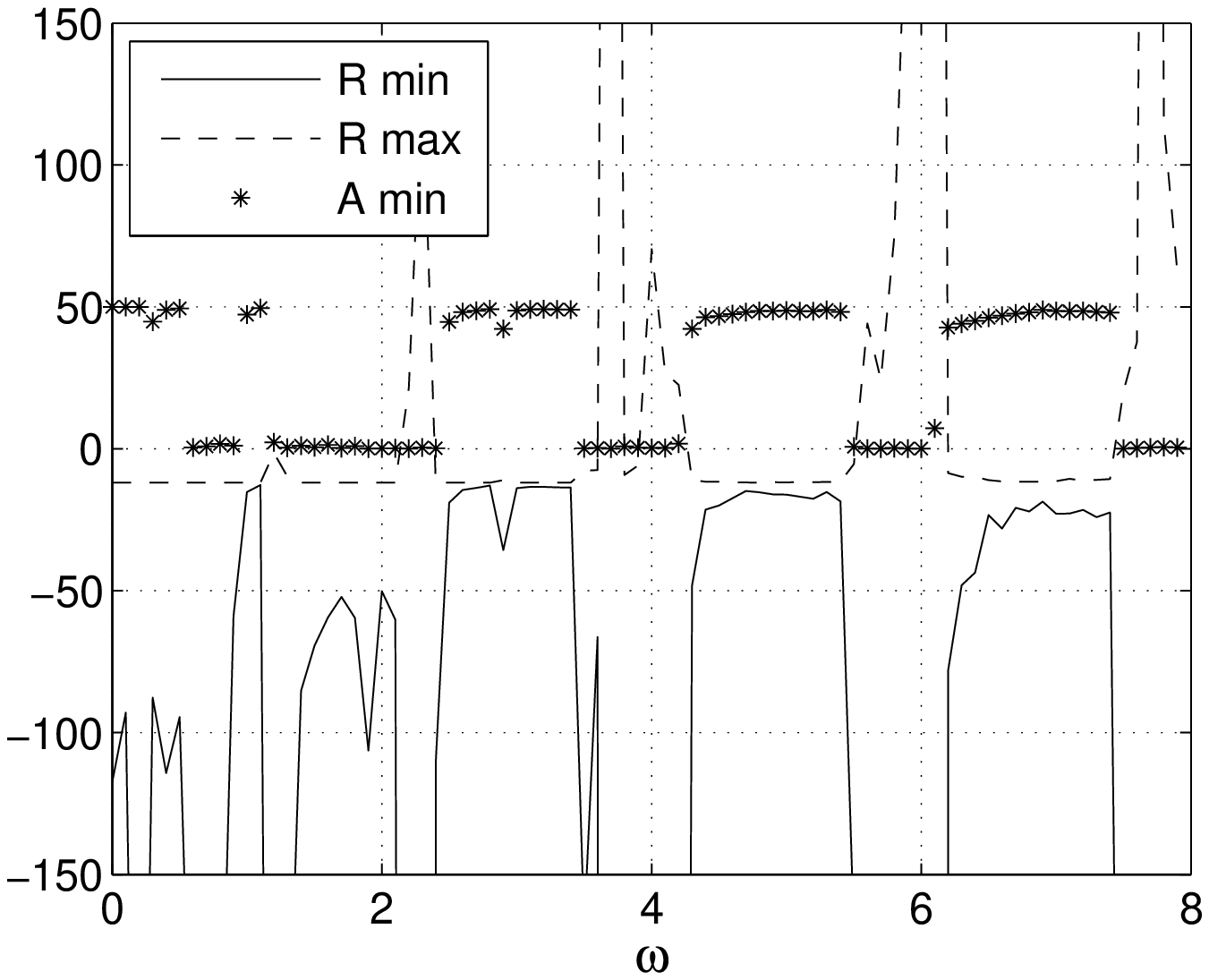}
\end{array}$
\end{center}
\caption{
Left panel: Maximum of the energy as function of $\omega$ for
$\phi_0=0.1$. 
Right panel: Maximum and minimum of the scalar curvature $R$ as
functions of $\omega$. 
The $*$s correspond to the minimum of $A$ in some units, showing the  
 correlation between the latter quantities and the formation of a
 horizon.} 
\label{summa-enecurva}
\end{figure}

\begin{figure}[!htp]
\centering{}
\includegraphics[width=0.6\linewidth]{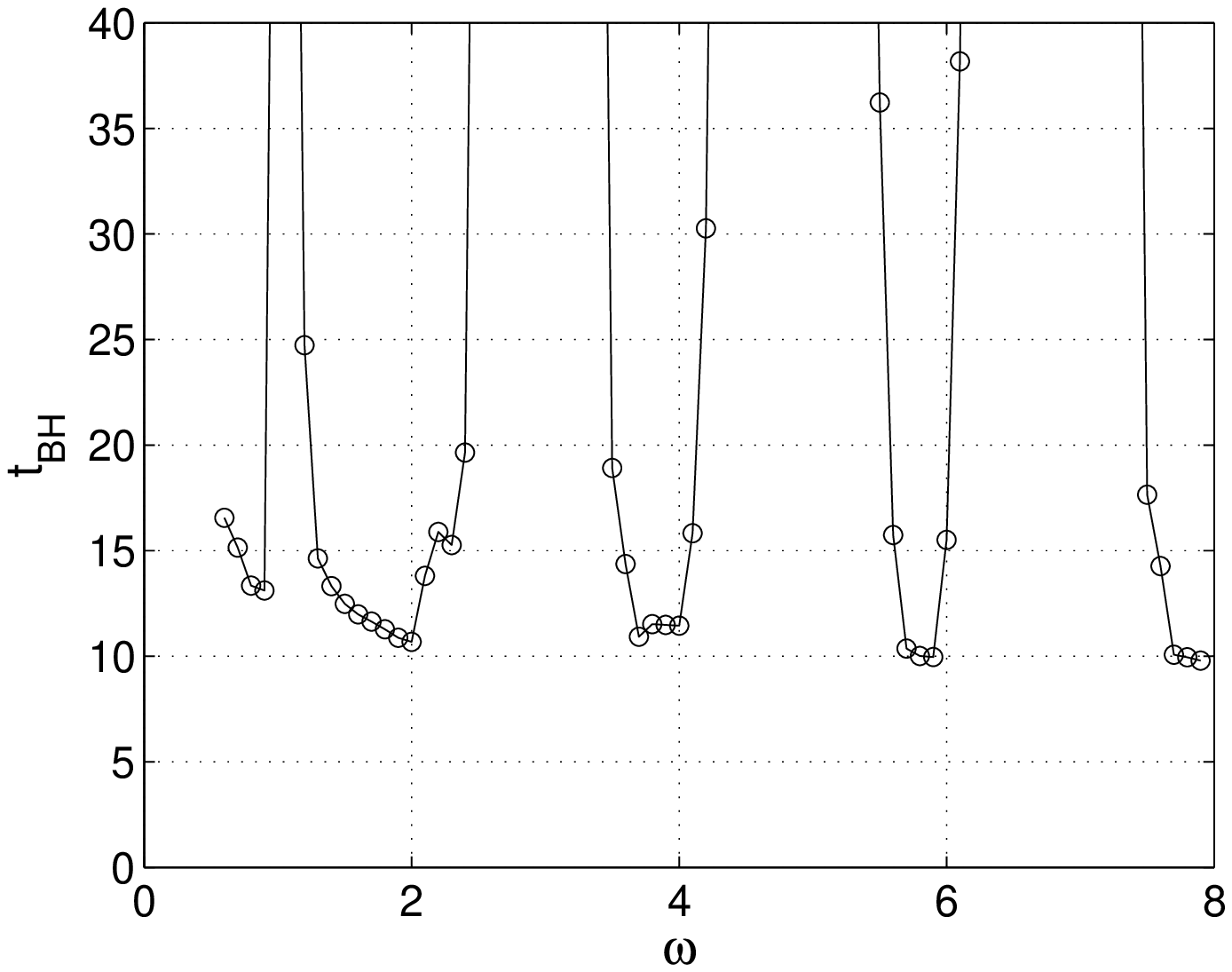}
\caption{The time of horizon formation as function of
$\omega$ for $\phi_0=0.1$.
In some intervals of frequency centered around $\omega$ taking on even
integer values, a BH horizon formation occurs rather quickly. 
On the other hand, we have observed no BH formation for odd integer
values, $\omega=3,5,7$, in calculations running up to the time
$t=800$. Horizon formation at $t=220$ is observed for $\omega=1$, 
while no BH formation is observed up to $t=800$ for $\omega=1.17$.} 
\label{summa-BH}
\end{figure}

For sufficiently large frequencies, we have checked that in the case
that no horizon has formed before $t_{\max}$, the maximum of the
energy and the maximum of $|R|$ remained small (see figure
\ref{summa-enecurva}). 
When a horizon forms, the scalar curvature $|R|$ tends to
become large; however this all happens inside of the horizon. 
The energy of the resulting BH does not diverge, on the contrary it
tends to become smaller for increasingly larger frequencies. 
In fig.~\ref{summa-BH} is shown the time of horizon formation 
 as function of $\omega$ for $\phi_0=0.1$. 

There are sharp transitions in the time of BH formation as a function
of $\omega$ (see fig.~\ref{summa-BH}). 
For frequencies $\omega$ around even integer values, a BH forms rather
quickly, whereas for frequencies $\omega$ near odd integer values, we
have observed no BH formation before stopping the numerical code at
$t=800$. 
A field propagating with the speed of light takes the time $\pi$ to
reach the boundary and return from $x=0$; this corresponds exactly to 
the frequency $\omega_0 = 2$. Perhaps this provides some qualitative
explanation for how come BHs form more easily subject to even
frequencies.

Two options are possible upon horizon formation. Either the solution
tends to a Schwarzschild BH at late times, as it is the only
time-independent solution compatible with our boundary conditions or
alternatively the BH will be time-dependent and its size and
mass will increase indefinitely -- we are effectively pumping energy
into the system. 
Unfortunately our choice of metric Ansatz \eqref{eq:bizonmetric} is
not permitting us to study this issue, because it becomes singular as 
soon as the horizon has formed. 

In the case of a massless scalar in flat 4-dimensional spacetime,
Choptuik found for an arbitrary family of solutions parametrized by
$p$, that both the critical horizon radius and the BH mass scale as 
$(p-p_*)^\gamma$, where $\gamma \approx 0.37$ and $p_*$ is the critical
value for BH formation \cite{Choptuik}. An identical scaling
was found for AdS${}_4$ with a massless scalar in
\cite{Husain:2002nk,Bizon}. 
For a massive field in flat space (e.g.~\cite{Brady:1997fj}),
depending on the class of solutions considered, it is also possible
that BH formation begins at finite mass and radius (type I
transition), instead of beginning at infinitesimal mass (type II
transition). In the case of a type II transition, the same critical
exponent $\gamma$ was found as in the massless case. 
The exponent $\gamma$ is not universal for all kinds of matter;
for example, in the case of type II transitions in Einstein-Yang-Mills
equations \cite{Choptuik:1996yg}, 
 $\gamma \approx 0.2$, (for a review of critical gravitational
collapse with various kinds of 
matter, see \cite{Gundlach:2007gc}). 

In AdS${}_4$ with a tachyonic scalar in the mass range
(\ref{massaranger}), it is in principle possible to study both the
critical behavior as function of the initial field profile, as well as 
function of the boundary condition at $x \rightarrow \pi/2$. 
The boundary condition (\ref{gradino}) 
gives rise to a type II transition, 
which we have observed due to the formation 
of arbitrarily small BH horizons with critical value 
$\phi_{0c} \approx 0.468$.
On the other hand, for the boundary conditions (\ref{perizo}) 
and fixed $\phi_0=0.1$, there
are several type I transitions as a function of $\omega$: for example, 
 BH formation begins at radius $ \approx 0.5$ and $t=64$ 
 for $\omega = 1.175 > \omega_c$,
 while no BH horizon has formed before $t=800$ for $\omega=1.17<\omega_c$.
We leave a more accurate study of these transitions and a
determination of the critical exponents in case of a type II
transition for future work.

\subsection{Numerical scheme}

We solve the time-dependent PDEs using the leapfrog method with a
five-point -- 4th order difference stencil for the spatial derivatives, 
a grid of $2500$ points and Courant number $0.5$.
Explicit dissipative terms are added.
The instanton solution (taken as the initial condition) is 
obtained by solving eqs.~(\ref{eq:inst1},\ref{eq:inst2})
using 5th order Runge-Kutta.
In order to initialize the leapfrog algorithm, we need to provide the
first and second order time derivatives of the initial instanton
profile function. We set all the first order time derivatives to zero,
except 
\beq
\dot{\Pi} = -\frac{\Psi e^{-\delta}} {\tan x} \, .
\eeq
The second derivatives read
\begin{align}
\ddot{\Psi}= - e^{-2 \delta} \left( \frac{V'(\phi)}{\cos x \sin x} +
\left(V(\phi)  -\frac{4}{\tan^2 x} \right) \Psi \right) \, , \qquad 
\ddot{\phi} = - \frac{\Psi A e^{-2 \delta}}{\tan x} \, , \non
\ddot{A}=A^2 e^{-2 \delta} \Psi^2 \cos^2 x \, ,
\end{align}
and $\ddot{\Pi}=0 $. 
No simple expression for $\ddot{\delta}$ is available and hence
$\delta$ is recalculated using eq.~(\ref{deltaeq}). 
For practical reasons we introduce an IR and a UV cutoff 
$x_{\rm min}=0.001$ and $x_{\rm max}=1.565$
(if we use values of $x_{\rm max}$  closer to $\pi/2$, 
the corresponding boundary conditions make the numerics unstable).
In order to update $\phi$ at the boundary $x_{\rm max}$
and $A$ at the origin $x_{\rm min}$ we use
eqs.~(\ref{tempoeq}). Subsequently eqs.~(\ref{Aeq},\ref{deltaeq}) 
are solved using a 5th order Runge-Kutta method. The boundary
value $\delta(x_{\rm max})$ is fixed to be zero.
The eqs.~(\ref{tempoeq}) are then used to estimate the numerical
accuracy.

\section{Discussion}

In this note we have discussed the issue of dynamical stability in
various examples of theories with  time-dependent potentials which
are periodically unbounded from below.
Considering an oscillating quadratic potential in the case of
classical and quantum mechanics as well as in field theory, the
stability regions are encoded by the phase diagram of the Mathieu
equation. The zero mode is stabilized when the frequency is high
enough. In addition we found a compactification scheme relying on
number theory such as to avoid resonances in all momentum modes of the
field theory.  
The effective Hamiltonian for periodically driven systems
\cite{Rahav:2003a,Rahav:2003b} suggests that a similar dynamical
stabilization occurs also for interacting theories. 
We have found numerical evidence for this in a theory with an AdS/CFT
dual; for sufficiently large $\omega$ the big-crunch singularity is
screened by a black hole horizon, preventing the crunch singularity to
reach the boundary in a finite time. 
We have observed sharp transitions in the black hole formation time as 
function of the frequency $\omega$. For frequencies near even integer
values, horizons form rather quickly while for those near odd integer
values, we have observed black hole formation at very late times or not at
all, due to finite simulation time. 

In fact we would be curious to see if such a potential could be a
holographic description of a universe oscillating between bangs and
crunches or healed bangs and healed crunches. An analytic study of
such system was carried out in
\cite{Nappi:1992kv,Kounnas:1992wc,Liu:2002ft,Elitzur:2002rt,Gasperini:2002bn} 
and references therein,
but there the system had an infinite amount of boundaries and  in the
presence of closed time-like curves. One could wish that all this
could be encapsulated on a single boundary without redundant
features. 
This could have been a system for which the time-dependent couplings
on the boundary are appropriate. Unfortunately, for this, the
numerical analysis is not a substitute for an exact solution. It is
worthwhile to study such systems non-perturbatively even if there is a  
lack of a strong phenomenological motivation at this stage.

Let us comment on possible topics of future directions:
\begin{itemize}

\item For the free field theory part one could 
consider different compactifications of the spatial 
dimensions to see if this gives rise to additional structure. 

\item 
It would be challenging to follow the black hole evolution after the
horizon has formed; for this it might be necessary to use a different
(non-polar) metric Ansatz and also introduce a singularity excision,
e.g.~as in \cite{Pretorius,Chesler:2008hg,Heller:2012je}. 

\item We did not address  here the precise nature of the 
transition between the big crunch and black hole of infinite mass;
for example, it would be interesting to determine the dependence 
of the black hole  mass near the critical frequency $\omega_c$ 
separating the black hole regime from the crunch.

\item A more detailed study of the transition type (I or II)
at the threshold of black hole formation, both as a function
of initial field profile $\phi_0$ and of the time-dependent boundary
condition $\omega$, would be worthwhile to investigate; in particular
calculating the critical exponents for the type II transition.

\end{itemize}

\acknowledgments

The authors thank Jos\'e Barb\'on, Shmuel Fishman, Romuald A.~Janik,
Barak Kol, Zohar Komargodski, Elon Lindenstrauss, Ioannis
Papadimitriou, Boris Pioline, Gabriele Veneziano and Shimon
Yankielovich for fruitful discussions. 
The work of S.~Elitzur is partially supported by 
the Israel Science Foundation Center of Excellence.
The work of R.~Auzzi, S.~B.~Gudnason and E.~Rabinovici is
partially supported by 
the American-Israeli Bi-National Science Foundation and
the Israel Science Foundation Center of Excellence.
S.~B.~Gudnason is also partially supported by the Golda Meir
Foundation Fund.

\appendix

\section{Estimating the characteristic exponent\label{app:nu}}

The characteristic exponent $\nu$ can be estimated numerically  by
calculating the solution $y(\tau)$ to eq.~\eqref{eq:Mathieu} subject
to the boundary conditions $y(0)=1$, $\dot{y}(0)=0$, and then
evaluated at $\tau=\pi$
\cite{Abramowitz} 
\beq
\cos(\nu\pi) = y(\pi) \ ,
\eeq
where $\nu$ is implicitly a function of $a,q$. 

\section{Estimating the amplitude\label{app:amplitude}}

From the interpretation that the generalized coordinate $x$ describes
a particle, a natural question would be how far the particle goes. 
Let us first consider the simple example of a particle at rest and
situated at position $x(0)=1$ at time $\tau=0$. Using an expansion
valid for small $q\ll 1$ and $\nu$ real but not an integer, the
solution reads \cite{Abramowitz} 
\begin{align}
C(\tau) = &\; \frac{
\cos[\nu\tau]
-\frac{q \cos\left[(\nu+2)\tau\right]}{4(\nu+1)}
+\frac{q\cos\left[(\nu-2)\tau\right]}{4(\nu-1)}
+\frac{q^2\cos\left[(\nu+4)\tau\right]}{32(\nu+1)(\nu+2)}
+\frac{q^2\cos\left[(\nu-4)\tau\right]}{32(\nu-1)(\nu-2)}
-\frac{q^2(\nu^2+1)\cos(\nu\tau)}{16(\nu^2-1)^2}}{1
-\frac{q}{4(\nu+1)}
+\frac{q}{4(\nu-1)}
+\frac{q^2}{32(\nu+1)(\nu+2)}
+\frac{q^2}{32(\nu-1)(\nu-2)}
-\frac{q^2(\nu^2+1)}{16(\nu^2-1)^2}} \non
& + \mathcal{O}(q^3) \, ,
\label{eq:mC}
\end{align}
which is called the Mathieu cosine function and from this we
can estimate the amplitude 
\begin{align}
\max(|C|) = &\;
1 + \frac{q}{\alpha} + \frac{\beta q^2}{8(\nu^2-1)^2} + \mathcal{O}(q^3) \, ,
\qquad \label{eq:mC_amplitude} \\
&\left\{
\begin{array}{ll}
\alpha = 1-\nu^2 \, , \beta = 5+\nu^2 \, , & {\rm for}\ \nu < 1 \, , \\
\alpha = 2+2\nu \, , \beta = \frac{-11+15\nu-9\nu^2+\nu^3}{2(\nu-2)} \, , & {\rm for}\ 1 < \nu < 2 \, , \\
\alpha = 2+2\nu \, , \beta = 3-2\nu+\nu^2 \, , & {\rm for}\ \nu > 2 \, .
\end{array}
\right. \nonumber
\end{align}
Let us first consider $a=0$. Using 
\beq
\cos(\nu\pi) \simeq \cos(\sqrt{a}\pi) 
+\frac{\pi q^2 \sin(\sqrt{a}\pi)}{4\sqrt{a}(a-1)} \, ,
\label{eq:cosnu_estimate}
\eeq
we obtain 
\beq
\nu \simeq \frac{q}{\sqrt{2}} + \mathcal{O}(q^3) \, , 
\eeq
which upon insertion in the amplitude \eqref{eq:mC_amplitude} gives 
\beq
\max(|C|) = 1+q+\frac{5q^2}{8} + \mathcal{O}(q^3) \, .
\eeq
For $a > 0$, the expansion of $\nu$ changes according to
\beq
\nu \simeq \sqrt{a} - (-1)^{[\sqrt{a}]}\frac{q^2}{4\sqrt{a}(a-1)} \, , 
\label{eq:nuexpansion}
\eeq
which to order $q^2$ then gives rise to the amplitude
\eqref{eq:mC_amplitude} with $\nu=\sqrt{a}$. 
The above amplitudes tell us how far the particle will go, i.e.~the 
absolute value of the largest value that the coordinate $|x(t)|$ will
take on, with the initial condition that the particle starts at rest
at $x(0)=1$ and time $\tau=0$, assuming that $(a,q)$ are in a
stability band of the phase diagram, see
fig.~\ref{fig:mathieu_phasediagram}. 
If one wishes to start with the particle situated at $x_0$, it
suffices to multiply the solution \eqref{eq:mC} by $x_0$ since the
equation of motion is linear and hence also the amplitude scales
accordingly.

A similar situation can be estimated as well, namely considering the
particle situated at the origin $x(0)=0$ but having velocity
$\dot{x}(0)=1$ at time $\tau = 0$
\begin{align}
S(\tau) = &\; \frac{
\sin[\nu\tau]
-\frac{q\sin\left[(\nu+2)\tau\right]}{4(\nu+1)}
+\frac{q\sin\left[(\nu-2)\tau\right]}{4(\nu-1)}
+\frac{q^2\sin\left[(\nu+4)\tau\right]}{32(\nu+1)(\nu+2)}
+\frac{q^2\sin\left[(\nu-4)\tau\right]}{32(\nu-1)(\nu-2)}
-\frac{q^2(\nu^2+1)\sin(\nu\tau)}{16(\nu^2-1)^2}}{1
-\frac{q(\nu+2)}{4(\nu+1)}
+\frac{q(\nu-2)}{4(\nu-1)}
+\frac{q^2(\nu+4)}{32(\nu+1)(\nu+2)}
+\frac{q^2(\nu-4)}{32(\nu-1)(\nu-2)}
-\frac{q^2(\nu^2+1)\nu}{16(\nu^2-1)^2}} \non
& + \mathcal{O}(q^3) \, ,
\label{eq:mS}
\end{align}
which is called the Mathieu sine function. From this
solution we can similarly estimate the amplitude of the particle's
movement as
\begin{align}
\max(|S|) = &\;
\frac{1}{\nu}\left(1 + \frac{\alpha q}{2(\nu-1)}
+ \frac{\beta q^2}{8(\nu^2-1)^2}\right) + \mathcal{O}(q^3) \,
, \label{eq:mS_amplitude} \\ 
&\left\{
\begin{array}{ll}
\alpha = 0 \, , \beta = \frac{(\nu^2-2)(\nu^2+5)}{\nu^2-4} \, , & {\rm
  for}\ \nu < 1 \, , \\ 
\alpha = 1 \, , \beta = \frac{17+15\nu+3\nu^2+\nu^3}{2(\nu+2)} \, , &
       {\rm for}\ 1 < \nu < 2 \, , \\ 
\alpha = 1 \, , \beta = \frac{-18-8\nu+5\nu^2+2\nu^3+\nu^4}{\nu^2-4}
\, , & {\rm for}\ \nu > 2 \, .
\end{array}
\right. \nonumber
\end{align}
For $a=0$, it can be expressed as
\beq
\max(|S|) = \frac{\sqrt{2}}{q} + \frac{5q}{8\sqrt{2}} -
\frac{5q^2}{16\sqrt{2}} + \mathcal{O}(q^3) \, ,
\eeq
while for $a>0$ we obtain
\begin{align}
\max(|S|) = &\; 
\frac{1}{\sqrt{a}}
+\frac{\alpha q}{2\sqrt{a}(\sqrt{a}-1)}
+\frac{\beta q^2}{8(a-1)^2 a^{3/2}} + \mathcal{O}(q^3) \, , \\
&\left\{
\begin{array}{ll}
\alpha = 0 \, , \beta = \frac{2(-1)^{[\sqrt{a}]}(a-4)(a-1) +
  a(a-2)(a+5)}{a-4} \, , & {\rm for}\ a < 1 \, , \\
\alpha = 1 \, , \beta =
\frac{4(-1)^{[\sqrt{a}]}\left(-2-\sqrt{a}+2a+a^{3/2}\right) +
  a\left(17+15\sqrt{a}+3a+a^{3/2}\right)}{2(\sqrt{a}+2)} \, , & {\rm 
  for}\ 1 < a < 4 \, , \\
\alpha = 1 \, , \beta = \frac{2(-1)^{[\sqrt{a}]}(a-4)(a-1) +
  a\left(-18-8\sqrt{a}+5a+2a^{3/2}+a^2\right)}{a-4} \, , & {\rm
  for}\ a < 4 \, , 
\end{array}
\right. \nonumber
\end{align}
where $[\sqrt{a}]$ denotes the integer part of $\sqrt{a}$. 
The above amplitudes tell us how far the particle will go, i.e.~the
absolute value of the largest value that the coordinate $|x(t)|$ will
take on, with the initial condition that the particle starts from
$x(0)=0$ with velocity $\dot{x}(0)=1$ at time $\tau=0$. 
If one wishes to start with the velocity $v_0$, it
suffices to multiply the solution \eqref{eq:mS} by $v_0$ since the 
equation of motion is linear and hence also the amplitude scales
accordingly.

\section{A diagrammatic derivation of the effective
  potential\label{app:FTderiv}}

Let us consider the classical mechanics problem
\beq
L = \frac{1}{2}m\dot{X}^2 - g X^n \cos(\omega t) \, ,
\eeq
with $n\geq 2$. Carrying out a Fourier transform of the field 
$X(t) = \int d\nu X(\nu) e^{i\nu t}$ and integrating
the Lagrangian over time to obtain the action, we get
\begin{align}
S =&\; \frac{1}{2}m \int d\nu \; \nu^2 X(\nu)X^*(\nu) \\
&-\frac{g}{2}\int d\nu_1 d\nu_2 \cdots d\nu_n \; X(\nu_1)X(\nu_2)\cdots X(\nu_n)
\delta(\nu_1+\nu_2+\cdots+\nu_n-\omega) \non 
&-\frac{g}{2}\int d\nu_1 d\nu_2 \cdots d\nu_n \; X(\nu_1)X(\nu_2)\cdots X(\nu_n)
\delta(\nu_1+\nu_2+\cdots+\nu_n+\omega) \, . \nonumber
\end{align}
To obtain the effective action, we will split the field $X(\nu)$ into
two fields according to whether $|\nu|<\Delta\ll \omega$ in which case we
will call the field still $X(\nu)$ or whether $|\nu|>\Delta$ where we
call it $\xi(\nu)$. This splits the field into the drift part $X$ and
the rapidly moving part $\xi$ and leads to the action
\begin{align}
S =&\; 
\frac{1}{2}m \int_{|\nu|<\Delta} d\nu \; \nu^2 X(\nu)X^*(\nu) 
+ \frac{1}{2}m \int_{|\nu|>\Delta} d\nu \; \nu^2 \xi(\nu)\xi^*(\nu)
\\
&- \frac{g}{2} \int_{|\nu_i|<\Delta}
d\nu_1 d\nu_2 \cdots d\nu_n \; X(\nu_1)X(\nu_2)\cdots X(\nu_n)
\delta(\nu_1+\nu_2+\cdots+\nu_n-\omega) \non
&- \frac{g}{2} \int_{|\nu_i|<\Delta}
d\nu_1 d\nu_2 \cdots d\nu_n \; X(\nu_1)X(\nu_2)\cdots X(\nu_n)
\delta(\nu_1+\nu_2+\cdots+\nu_n+\omega) \non
&- \frac{g n}{2} \int_{\begin{subarray}{l}|\nu_1|>\Delta\\|\nu_i|<\Delta\end{subarray}} d\nu_1 d\nu_2 \cdots
d\nu_n \; \xi(\nu_1)X(\nu_2)\cdots X(\nu_n)
\delta(\nu_1+\nu_2+\cdots+\nu_n-\omega) \non 
&- \frac{g n}{2} \int_{\begin{subarray}{l}|\nu_1|>\Delta\\|\nu_i|<\Delta\end{subarray}} d\nu_1 d\nu_2 \cdots
d\nu_n \; \xi(\nu_1)X(\nu_2)\cdots X(\nu_n)
\delta(\nu_1+\nu_2+\cdots+\nu_n+\omega) \non
&- \frac{g n(n-1)}{4}
\int_{\begin{subarray}{l}|\nu_1|>\Delta\\|\nu_2|>\Delta\\|\nu_i|<\Delta\end{subarray}}
d\nu_1 d\nu_2 \cdots d\nu_n \; \xi(\nu_1)\xi(\nu_2)X(\nu_3)\cdots X(\nu_n)
\delta(\nu_1+\nu_2+\cdots+\nu_n-\omega) \non
&- \frac{g n(n-1)}{4}
\int_{\begin{subarray}{l}|\nu_1|>\Delta\\|\nu_2|>\Delta\\|\nu_i|<\Delta\end{subarray}}
d\nu_1 d\nu_2 \cdots d\nu_n \; \xi(\nu_1)\xi(\nu_2)X(\nu_3)\cdots X(\nu_n)
\delta(\nu_1+\nu_2+\cdots+\nu_n+\omega) \non
&\phantom{-\ }\vdots\non
&- \frac{g}{2} \int_{|\nu_i|>\Delta}
d\nu_1 d\nu_2 \cdots d\nu_n \; \xi(\nu_1)\xi(\nu_2)\cdots \xi(\nu_n)
\delta(\nu_1+\nu_2+\cdots+\nu_n-\omega) \non
&- \frac{g}{2} \int_{|\nu_i|>\Delta}
d\nu_1 d\nu_2 \cdots d\nu_n \; \xi(\nu_1)\xi(\nu_2)\cdots \xi(\nu_n)
\delta(\nu_1+\nu_2+\cdots+\nu_n+\omega) \, . \nonumber
\end{align}
To get the effective action one solves $\xi$ in terms of $X$ and
substitutes back into the action. Typically the energy $\nu$ of
$\xi(\nu)$ is of order $\omega$ so $\xi$ is of order $1/\omega^2$.
The solution of $\xi$ in terms of $X$ can then be done iteratively and
be presented by diagrams. 
From this action we can read off the propagator as well as the
vertices for calculating Feynman diagrams in order to construct the
effective action in terms of the field $X(\nu)$. 
Let us note that the second and third lines in the above action tell
us that to zeroth order in $\xi$ there is no interaction among the drift
part of the field $X$ as the sum of $\nu_1+\nu_2+\cdots+\nu_n$ is also
much smaller than $\omega$ unless $n$ is a huge integer. At this point
we have established for finite (and reasonable) $n$, the effective
action has no potential to order $g$. Let us now calculate the leading
correction (at order $g^2$) to the effective potential case by case.

\subsection{Quadratic case}

For $n=2$ there is a subtlety that there is some sort of mass
term. We will not sum up the mass insertions but instead consider them
as interactions and hence the propagator of $\xi$ is
\beq
D = -\frac{i}{m\nu^2} \, ,
\label{eq:xiprop}
\eeq
while the leading vertices are
\beq
\oplus = - i g \delta(\nu_1+\nu_2+\omega)\, , \qquad
\ominus = - i g \delta(\nu_1+\nu_2-\omega)\, . 
\eeq
To order $g^2$ there are two tree-level diagrams
\setlength{\unitlength}{1pt}
\begin{center}
\begin{picture}(120,30)
\drawline(0,15)(40,15)
\drawline(80,15)(120,15)
\begin{dashjoin}[4]{5}
\jput(40,15){\cb}\jput(80,15){\cb}
\end{dashjoin}
\put(40,20){\makebox(0,0)[b]{$\ominus$}}
\put(80,20){\makebox(0,0)[b]{$\oplus$}}
\put(20,0){\makebox(0,0)[b]{\small $X(\nu)$}}
\put(60,0){\makebox(0,0)[b]{\small $\xi(\nu-\omega)$}}
\put(100,0){\makebox(0,0)[b]{\small $X(\nu)$}}
\end{picture}\quad
\raisebox{2.4ex}{$+$}\quad
\begin{picture}(120,30)
\drawline(0,15)(40,15)
\drawline(80,15)(120,15)
\begin{dashjoin}[4]{5}
\jput(40,15){\cb}\jput(80,15){\cb}
\end{dashjoin}
\put(40,20){\makebox(0,0)[b]{$\oplus$}}
\put(80,20){\makebox(0,0)[b]{$\ominus$}}
\put(20,0){\makebox(0,0)[b]{\small $X(\nu)$}}
\put(60,0){\makebox(0,0)[b]{\small $\xi(\nu+\omega)$}}
\put(100,0){\makebox(0,0)[b]{\small $X(\nu)$}}
\end{picture}
\end{center}
which gives the amplitude when taking into account the appropriate
symmetry factor
\beq
\frac{g^2}{2m(\omega-\nu)^2} X^2(\nu) 
+ \frac{g^2}{2m(\omega+\nu)^2} X^2(\nu) = 
\frac{g^2}{m}\left[\frac{1}{\omega^2} +
\frac{3\nu^2}{\omega^4} + \mathcal{O}(\omega^{-6})\right] X^2(\nu) \, .
\eeq
To obtain the effective potential we need to make the inverse Fourier
transform, such that $\nu X$ becomes $\dot{X}$. Finally we get
\beq
V_{\rm eff} = \frac{g^2}{m\omega^2} X^2 
+ \frac{3g^2}{m\omega^4} \dot{X}^2 \, ,
\eeq 
which equals that of eq.~\eqref{eq:Heffx^2} when identifying
$g=\lambda^3$.

\subsection{Quartic case}

The $n=4$ case is analogous but involves slightly more combinatorics. 
The propagator of $\xi$ is still that of eq.~\eqref{eq:xiprop} and the
leading vertices are 
\beq
\oplus = - i 12 g \delta(\nu_1+\nu_2+\nu_3+\nu_4+\omega)\, , \qquad
\ominus = - i 12 g \delta(\nu_1+\nu_2+\nu_3+\nu_4-\omega)\, . 
\eeq
To order $g^2$ there are two tree-level diagrams:
\setlength{\unitlength}{1pt}
\begin{center}
\begin{picture}(240,60)
\drawline(10,60)(60,30)
\drawline(0,30)(60,30)
\drawline(10,0)(60,30)
\drawline(180,30)(230,60)
\drawline(180,30)(240,30)
\drawline(180,30)(230,0)
\begin{dashjoin}[4]{5}
\jput(60,30){\cb}\jput(180,30){\cb}
\end{dashjoin}
\put(60,40){\makebox(0,0)[b]{$\ominus$}}
\put(180,40){\makebox(0,0)[b]{$\oplus$}}
\put(35,55){\makebox(0,0)[b]{\small $X(\nu_1)$}}
\put(20,30){\makebox(0,0)[b]{\small $X(\nu_2)$}}
\put(45,0){\makebox(0,0)[b]{\small $X(\nu_3)$}}
\put(120,15){\makebox(0,0)[b]{\small $\xi(\nu_1+\nu_2+\nu_3-\omega)$}}
\put(205,55){\makebox(0,0)[b]{\small $X(\nu_1')$}}
\put(220,30){\makebox(0,0)[b]{\small $X(\nu_2')$}}
\put(195,0){\makebox(0,0)[b]{\small $X(\nu_3')$}}
\end{picture}
\end{center}
and the same with $\ominus$ and $\oplus$ interchanged. 
Taking into account the appropriate symmetry factor, the sum of the
above diagrams gives the amplitude
\begin{align}
&\left[\frac{2g^2}{m(\omega-\nu_1-\nu_2-\nu_3)^2}
+\frac{2g^2}{m(\omega+\nu_1+\nu_2+\nu_3)^2} 
\right]\times \non
&\qquad X(\nu_1)X(\nu_2)X(\nu_3)X(\nu_1')X(\nu_2')X(\nu_3')
\delta(\nu_1+\nu_2+\nu_3-\nu_1'-\nu_2'-\nu_3') \non
&=\frac{4g^2}{m} 
\left[\frac{1}{\omega^2}+\frac{3(\nu_1+\nu_2+\nu_3)^2}{\omega^4} +
\mathcal{O}(\omega^{-6})\right] \times \non 
&\qquad X(\nu_1)X(\nu_2)X(\nu_3)
X(\nu_1')X(\nu_2')X(\nu_3')\delta(\nu_1+\nu_2+\nu_3-\nu_1'-\nu_2'-\nu_3')
\, .
\end{align}
As before, to obtain the effective potential we need to make the
inverse Fourier transform, such that $\nu X$ becomes
$\dot{X}$. Finally we get 
\beq
V_{\rm eff} = \frac{4g^2}{m\omega^2} X^6
+ \frac{108g^2}{m\omega^4} X^4 \dot{X}^2 \, ,
\eeq 
which equals that of \eqref{eq:Heffx^4} when identifying
$g=\lambda^5$.

The above effective potentials are conservative and independent of
time even though the original Lagrangian was time-dependent. This
effective conservation of energy can be understood as follows. By
means of our approximation, we have $|\nu_i|\ll\omega$. If we look at
the above tree-level diagram we can see that by starting with a
$\ominus$ vertex, the only way to obey the constraint that $|\nu_i'|$
be $\ll\omega$ is to end the diagram with a $\oplus$ vertex. We can
construct bigger tree diagrams (to higher order in $g$ and in
$1/\omega$) by inserting for instance a vertex with two $X$s and two 
$\xi$s (which comes with a $g$). As long as the number of inserted
vertices $k$ is small enough we need the same number of $\ominus$ as
$\oplus$ vertices
\setlength{\unitlength}{1pt}
\begin{center}
\begin{picture}(240,40)
\drawline(5,40)(30,20)
\drawline(0,20)(30,20)
\drawline(5,0)(30,20)
\drawline(210,20)(235,40)
\drawline(210,20)(240,20)
\drawline(210,20)(235,0)
\begin{dashjoin}[4]{5}
\jput(30,20){\cb}\jput(60,20){\cb}\jput(90,20){}
\end{dashjoin}
\begin{dashjoin}[4]{5}
\jput(210,20){\cb}\jput(180,20){\cb}\jput(150,20){}
\end{dashjoin}
\drawline(45,0)(60,20)
\drawline(75,0)(60,20)
\drawline(165,0)(180,20)
\drawline(195,0)(180,20)
\put(30,30){\makebox(0,0)[b]{$\ominus$}}
\put(60,30){\makebox(0,0)[b]{$\oplus$}}
\put(60,5){\makebox(0,0)[b]{\small $1$}}
\put(120,20){\makebox(0,0)[b]{$\ldots$}}
\put(180,30){\makebox(0,0)[b]{$\ominus$}}
\put(180,5){\makebox(0,0)[b]{\small $k$}}
\put(210,30){\makebox(0,0)[b]{$\oplus$}} 
\end{picture}
\end{center}
Hence, to this order the effective action conserves the energy of the
drift degrees of freedom $X$ and does not have explicit time
dependence. 
This diagram is just an example, many other tree-level diagrams of
different shape can be constructed. 
We can now estimate for how large $k$ we will be able to end the
diagram with $\ominus$ vertex and hence introduce $\omega$ dependence
in the effective action. Generically, when the number of $\ominus$
vertices is different from the number of $\oplus$ vertices, the
effective action re-acquires time dependence.  
If $\omega/\Delta$ is of order of the number of external legs
$L=2(3+k)$ which corresponds to order $\omega^{-2(k+1)}$ and
$g^{2+k}$, that is to order $\omega/\Delta-4$ in
$1/\omega$, \footnote{$\Delta$ here should be taken a factor of a few
  larger than the eigenfrequencies of the drift part of the system. }
then the effective potential will not be conservative. 
The argument holds for any diagram and not only the presented
example.


\begin{thebibliography}{99}

\bibitem{CdL}
  S.~R.~Coleman and F.~De Luccia,
  ``Gravitational Effects on and of Vacuum Decay,''
  Phys.\ Rev.\ D {\bf 21} (1980) 3305.

\bibitem{HH1}
 T.~Hertog and G.~T.~Horowitz,
  ``Towards a big crunch dual,''
  JHEP {\bf 0407} (2004) 073
  [hep-th/0406134].

\bibitem{HH2}
  T.~Hertog and G.~T.~Horowitz,
  ``Holographic description of AdS cosmologies,''
  JHEP {\bf 0504} (2005) 005
  [hep-th/0503071].

\bibitem{Elitzur:2005kz}
  S.~Elitzur, A.~Giveon, M.~Porrati and E.~Rabinovici,
  ``Multitrace deformations of vector and adjoint theories and their holographic duals,''
  JHEP {\bf 0602} (2006) 006
  [hep-th/0511061].

\bibitem{Elitzur:2007zz} 
  S.~Elitzur, A.~Giveon, M.~Porrati and E.~Rabinovici,
  ``Multitrace deformations of vector and adjoint theories and their holographic duals,''
  Nucl.\ Phys.\ Proc.\ Suppl.\  {\bf 171}, 231 (2007).

\bibitem{Craps}
  B.~Craps, T.~Hertog and N.~Turok,
  ``Quantum Resolution of Cosmological Singularities using AdS/CFT,''
  arXiv:0712.4180 [hep-th].

\bibitem{Craps:2009qc}
  B.~Craps, T.~Hertog and N.~Turok,
  ``A Multitrace deformation of ABJM theory,''
  Phys.\ Rev.\ D {\bf 80} (2009) 086007
  [arXiv:0905.0709 [hep-th]].
  
\bibitem{Bernamonti:2009dp}
  A.~Bernamonti and B.~Craps,
  ``D-Brane Potentials from Multi-Trace Deformations in AdS/CFT,''
  JHEP {\bf 0908} (2009) 112
  [arXiv:0907.0889 [hep-th]].

\bibitem{de Haro:2006nv}
  S.~de Haro, I.~Papadimitriou and A.~C.~Petkou,
  ``Conformally Coupled Scalars, Instantons and Vacuum Instability in AdS(4),''
  Phys.\ Rev.\ Lett.\  {\bf 98} (2007) 231601
  [hep-th/0611315].

\bibitem{Barbon:2010gn}
  J.~L.~F.~Barbon and E.~Rabinovici,
  ``Holography of AdS vacuum bubbles,''
  JHEP {\bf 1004} (2010) 123
  [arXiv:1003.4966 [hep-th]].

\bibitem{Harlow:2010az} 
  D.~Harlow,
  ``Metastability in Anti de Sitter Space,''
  arXiv:1003.5909 [hep-th].

\bibitem{Maldacena:2010un}
  J.~Maldacena,
  ``Vacuum decay into Anti de Sitter space,''
  arXiv:1012.0274 [hep-th].
  
\bibitem{Barbon:2011ta}
  J.~L.~F.~Barbon and E.~Rabinovici,
  ``AdS Crunches, CFT Falls And Cosmological Complementarity,''
  JHEP {\bf 1104} (2011) 044
  [arXiv:1102.3015 [hep-th]].

\bibitem{Kapitza}
  P.~L.~Kapitza,
  ``Dynamical stability of a pendulum when its point of suspension
  vibrates,'' 
  Zhur.\ Eksp.\ i\ Teoret.\ Fiz.\ {\bf 21}, 588 (1951);
  Collected Papers, ch.\ 45, p.\ 714.

\bibitem{Traschen:1990sw} 
  J.~H.~Traschen and R.~H.~Brandenberger,
  ``Particle Production During Out-of-equilibrium Phase Transitions,''
  Phys.\ Rev.\ D {\bf 42}, 2491 (1990).

\bibitem{Dolgov:1989us} 
  A.~D.~Dolgov and D.~P.~Kirilova,
  ``On Particle Creation By A Time Dependent Scalar Field,''
  Sov.\ J.\ Nucl.\ Phys.\  {\bf 51}, 172 (1990)
  [Yad.\ Fiz.\  {\bf 51}, 273 (1990)].

\bibitem{Kofman:1994rk}
  L.~Kofman, A.~D.~Linde and A.~A.~Starobinsky,
  ``Reheating after inflation,''
  Phys.\ Rev.\ Lett.\  {\bf 73} (1994) 3195
  [hep-th/9405187].

\bibitem{Kofman:1997yn}
  L.~Kofman, A.~D.~Linde and A.~A.~Starobinsky,
  ``Towards the theory of reheating after inflation,''
  Phys.\ Rev.\ D {\bf 56} (1997) 3258
  [hep-ph/9704452].

\bibitem{Durin:2003gj} 
  B.~Durin and B.~Pioline,
  ``Open strings in relativistic ion traps,''
  JHEP {\bf 0305}, 035 (2003)
  [hep-th/0302159].

\bibitem{Rahav:2003c}
  I.~Gilary, N.~Moiseyev, S.~Rahav and S.~Fishman, 
  ``Trapping of particles by lasers: the quantum Kapitza pendulum,''
  J.\ Phys.\ A{\bf 36}, (2003) L406-L415.

\bibitem{Pretorius}
  F.~Pretorius and M.~W.~Choptuik,
  ``Gravitational collapse in (2+1)-dimensional AdS space-time,''
  Phys.\ Rev.\ D {\bf 62} (2000) 124012
  [gr-qc/0007008].

\bibitem{Bizon}
  P.~Bizon and A.~Rostworowski,
  ``On weakly turbulent instability of anti-de Sitter space,''
  Phys.\ Rev.\ Lett.\  {\bf 107} (2011) 031102
  [arXiv:1104.3702 [gr-qc]].
 
 \bibitem{Bizon2}
  J.~Jalmuzna, A.~Rostworowski and P.~Bizon,
  ``A Comment on AdS collapse of a scalar field in higher dimensions,''
  Phys.\ Rev.\ D {\bf 84} (2011) 085021
  [arXiv:1108.4539 [gr-qc]].
 
\bibitem{GPZ}
   D.~Garfinkle and L.~A.~Pando Zayas,
  ``Rapid Thermalization in Field Theory from Gravitational Collapse,''
  Phys.\ Rev.\ D {\bf 84} (2011) 066006
  [arXiv:1106.2339 [hep-th]].

\bibitem{GPZ2}
  D.~Garfinkle, L.~A.~Pando Zayas and D.~Reichmann,
  ``On Field Theory Thermalization from Gravitational Collapse,''
  JHEP {\bf 1202} (2012) 119
  [arXiv:1110.5823 [hep-th]].

\bibitem{Witten:2001ua}
  E.~Witten,
  ``Multitrace operators, boundary conditions, and AdS / CFT correspondence,''
  hep-th/0112258.

\bibitem{BSS}
  M.~Berkooz, A.~Sever and A.~Shomer,
  ``'Double trace' deformations, boundary conditions and space-time singularities,''
  JHEP {\bf 0205} (2002) 034
  [hep-th/0112264].
  
\bibitem{SS}
  A.~Sever and A.~Shomer,
  ``A Note on multitrace deformations and AdS/CFT,''
  JHEP {\bf 0207} (2002) 027
  [hep-th/0203168].

\bibitem{KW}
  I.~R.~Klebanov and E.~Witten,
  ``AdS / CFT correspondence and symmetry breaking,''
  Nucl.\ Phys.\ B {\bf 556} (1999) 89
  [hep-th/9905104].

\bibitem{LL}
  L.~D.~Landau, and E.~M.~Lifshitz,
  ``Mechanics,''
  Pergamon Press (1960). 

\bibitem{Rahav:2003a}
  S.~Rahav, I.~Gilary, and S.~Fishman,
  ``Effective Hamiltonians for periodically driven systems,''
  Phys.\ Rev.\ {\bf A68}, (2003) 013820.
  [nlin/0301033].

\bibitem{Rahav:2003b}
  S.~Rahav, I.~Gilary, and S.~Fishman,
  ``Time independent description of rapidly oscillating potentials,''
  Phys.\ Rev.\ Lett.\ {\bf 91}, (2003) 110404.
  [nlin/0301033].

\bibitem{Rahav:2004}
  S.~Rahav, E.~Geva, and S.~Fishman,
  ``Time-independent approximations for periodically driven systems
  with friction,''
  Phys.\ Rev.\ E {\bf 71} (2005) 036210.
  [nlin/0408030].

\bibitem{Abramowitz}
  M.~Abramowitz and I.~A.~Stegun, 
  ``Handbook of Mathematical Functions'', 
  National Bureau of Standard, Applied Mathematics Series 55, Tenth
  edition, (1972).

\bibitem{Grozdanov:1988}
  T.~P.~Grozdanov, and M.~J.~Rakovi\'c,
  ``Quantum system driven by rapidly varying periodic perturbation,''
  Phys.\ Rev.\ {\bf A38}, 1739 (1988).

\bibitem{Lewis:1968tm}
  H.~R.~Lewis and W.~B.~Riesenfeld,
  ``An Exact quantum theory of the time dependent harmonic oscillator and of a charged particle time dependent electromagnetic field,''
  J.\ Math.\ Phys.\  {\bf 10} (1969) 1458.

\bibitem{Lewis:1967}
  H.~R.~Lewis,
  ``Classical and Quantum Systems with Time-dependent
  Harmonic-oscillator-type Hamiltonians,''
  Phys.\ Rev.\ Lett.\ {\bf 18}, 510-512 (1967). 

\bibitem{Pinney:1950}
  E.~Pinney,
  ``The nonlinear differential equation $y''(x) + p(x)y+cy^{-3}=0$,''
  Proceedings of the American Mathematical Society, {\bf 1}, 681
  (1950). 

\bibitem{Leach:2008}
  P.~G.~L.~Leach, and K.~Andriopoulos,
  ``The Ermakov Equation: A Commentary,''
  Appl.\ Anal.\ Discrete\ Math.\ {\bf 2}, 146-157 (2008).

\bibitem{Mathieu}
  E.~Mathieu,
  ``M\'emoire sur Le Mouvement Vibratoire d’une Membrane de forme
  Elliptique,''
  Journal des Math\'ematiques Pures et Appliqu\'ees, 137–203 (1868).

\bibitem{Lang}
   S. Lang, 
  ``An introduction to Diophantine approximations,'' 
  Addison-Wesley Pub. Co. (1966).

\bibitem{Cassels}
  J. W. S. Cassels, 
  ``An introduction to Diophantine approximation,''
  Haffner Pub.~Co., NY (1957, reprinted 1972).

\bibitem{McLachlan}
  N.~W.~McLachlan,
  ``Theory and Application of Mathieu Functions,''
  Oxford University Press (1951). 

\bibitem{Vergel:2009st}
  D.~Gomez Vergel and E.~J.~S.~Villasenor,
  ``The Time-dependent quantum harmonic oscillator revisited: Applications to Quantum Field Theory,''
  Annals Phys.\  {\bf 324} (2009) 1360
  [arXiv:0903.0289 [math-ph]].

\bibitem{ABJM}
  O.~Aharony, O.~Bergman, D.~L.~Jafferis and J.~Maldacena,
  ``N=6 superconformal Chern-Simons-matter theories, M2-branes and their gravity duals,''
  JHEP {\bf 0810} (2008) 091
  [arXiv:0806.1218 [hep-th]].

\bibitem{Duff:1999gh}
  M.~J.~Duff and J.~T.~Liu,
  ``Anti-de Sitter black holes in gauged N = 8 supergravity,''
  Nucl.\ Phys.\ B {\bf 554} (1999) 237
  [hep-th/9901149].

\bibitem{BF}
  P.~Breitenlohner and D.~Z.~Freedman,
  ``Stability in Gauged Extended Supergravity,''
  Annals Phys.\  {\bf 144} (1982) 249.

\bibitem{Hertog:2004dr}
  T.~Hertog and K.~Maeda,
  ``Black holes with scalar hair and asymptotics in N = 8 supergravity,''
  JHEP {\bf 0407} (2004) 051
  [hep-th/0404261].

\bibitem{Balasubramanian:1999re}
  V.~Balasubramanian and P.~Kraus,
  ``A Stress tensor for Anti-de Sitter gravity,''
  Commun.\ Math.\ Phys.\  {\bf 208} (1999) 413
  [hep-th/9902121].

\bibitem{Papadimitriou:2007sj}
  I.~Papadimitriou,
  ``Multi-Trace Deformations in AdS/CFT: Exploring the Vacuum Structure of the Deformed CFT,''
  JHEP {\bf 0705} (2007) 075
  [hep-th/0703152].

\bibitem{Hertog:2004ns}
  T.~Hertog and G.~T.~Horowitz,
  ``Designer gravity and field theory effective potentials,''
  Phys.\ Rev.\ Lett.\  {\bf 94} (2005) 221301
  [hep-th/0412169].

\bibitem{book}
  T.~W.~Baumgarte, and S.~L.~Shapiro, 
  ``Numerical Relativity: Solving Einstein's Equations on the Computer,''
  Cambridge University Press.

\bibitem{lectures}
  Matthew W.~Choptuik,
 ``Numerical Analysis with Applications in Theoretical Physics,''
  Lectures for Taller de Verano 1999 de FENOMEC. 

\bibitem{Dias}
  O.~J.~C.~Dias, G.~T.~Horowitz and J.~E.~Santos,
  ``Gravitational Turbulent Instability of Anti-de Sitter Space,''
  arXiv:1109.1825 [hep-th].

\bibitem{deOliveira:2012ac}
  H.~P.~de Oliveira, L.~A.~P.~Zayas and C.~A.~Terrero-Escalante,
  ``Turbulence and Chaos in Anti-de-Sitter Gravity,''
  arXiv:1205.3232 [hep-th].

\bibitem{Choptuik}
  M.~W.~Choptuik,
  ``Universality and scaling in gravitational collapse of a massless scalar field,''
  Phys.\ Rev.\ Lett.\  {\bf 70} (1993) 9.

\bibitem{Husain:2002nk}
  V.~Husain, G.~Kunstatter, B.~Preston and M.~Birukou,
  ``Anti-de Sitter gravitational collapse,''
  Class.\ Quant.\ Grav.\  {\bf 20} (2003) L23
  [gr-qc/0210011].

\bibitem{Brady:1997fj}
  P.~R.~Brady, C.~M.~Chambers and S.~M.~C.~V.~Goncalves,
  ``Phases of massive scalar field collapse,''
  Phys.\ Rev.\ D {\bf 56} (1997) 6057
  [gr-qc/9709014].

\bibitem{Choptuik:1996yg}
  M.~W.~Choptuik, T.~Chmaj and P.~Bizon,
  ``Critical behavior in gravitational collapse of a Yang-Mills field,''
  Phys.\ Rev.\ Lett.\  {\bf 77} (1996) 424
  [gr-qc/9603051].

\bibitem{Gundlach:2007gc}
  C.~Gundlach and J.~M.~Martin-Garcia,
  ``Critical phenomena in gravitational collapse,''
  Living Rev.\ Rel.\  {\bf 10} (2007) 5
  [arXiv:0711.4620 [gr-qc]].

\bibitem{Nappi:1992kv} 
  C.~R.~Nappi and E.~Witten,
  ``A Closed, expanding universe in string theory,''
  Phys.\ Lett.\ B {\bf 293}, 309 (1992)
  [hep-th/9206078].

\bibitem{Kounnas:1992wc} 
  C.~Kounnas and D.~Lust,
  ``Cosmological string backgrounds from gauged WZW models,''
  Phys.\ Lett.\ B {\bf 289}, 56 (1992)
  [hep-th/9205046].

\bibitem{Liu:2002ft}
  H.~Liu, G.~W.~Moore and N.~Seiberg,
  ``Strings in a time dependent orbifold,''
  JHEP {\bf 0206} (2002) 045
  [hep-th/0204168].

\bibitem{Elitzur:2002rt} 
  S.~Elitzur, A.~Giveon, D.~Kutasov and E.~Rabinovici,
  ``From big bang to big crunch and beyond,''
  JHEP {\bf 0206}, 017 (2002)
  [hep-th/0204189].

\bibitem{Gasperini:2002bn} 
  M.~Gasperini and G.~Veneziano,
  ``The Pre - big bang scenario in string cosmology,''
  Phys.\ Rept.\  {\bf 373}, 1 (2003)
  [hep-th/0207130].

\bibitem{Heller:2012je}
  M.~P.~Heller, R.~A.~Janik and P.~Witaszczyk,
  ``A numerical relativity approach to the initial value problem in asymptotically Anti-de Sitter spacetime for plasma thermalization - an ADM formulation,''
  arXiv:1203.0755 [hep-th].

\bibitem{Chesler:2008hg}
  P.~M.~Chesler and L.~G.~Yaffe,
  ``Horizon formation and far-from-equilibrium isotropization in supersymmetric Yang-Mills plasma,''
  Phys.\ Rev.\ Lett.\  {\bf 102} (2009) 211601
  [arXiv:0812.2053 [hep-th]].

\end{thebibliography}
\end{document}